\documentclass[a4paper,fleqn,usenatbib]{mnras}

\usepackage{newtxtext,newtxmath}

\usepackage[T1]{fontenc}
\usepackage{ae,aecompl}

\usepackage{graphicx}	
\usepackage{amsmath}	

\title[Compositional Turbulence and Layering]{Compositional Turbulence and Layering in the Gaseous Envelopes of Forming Planets}

\author[K. Menou \& H. T. Zhang]{
  Kristen Menou$^{1, 2, 3}$
\& Hong Tao Zhang$^{4}$  
\\
$^{1}$  Physics \& Astrophysics Group, Dept.   of  Physical  \&  Environmental  Sciences,  University  of  Toronto  Scarborough,\\   1265 Military Trail, Toronto, Ontario, M1C 1A4, Canada \\
$^{2}$ David A. Dunlap Department  of Astronomy \& Astrophysics, University of Toronto.
50 St.  George Street, Toronto, Ontario, M5S 3H4, Canada 
\\
$^{3}$ Department of Physics, University of Toronto,
60 St George Street, Toronto, Ontario, M5S 1A7, Canada \\
$^{4}$ Department of Mathematics,  University of Toronto,  Bahen Centre, 40 St. George St. M5S 2E4, Canada
}

\date{Accepted XXX. Received YYY; in original form ZZZ}

\begin{document}
\label{firstpage}
\pagerange{\pageref{firstpage}--\pageref{lastpage}}
\maketitle

\begin{abstract}
Differential settling and growth of dust grains impact the structure of the radiative envelopes of gaseous planets during formation.  Sufficiently rapid dust growth can result in envelopes with substantially reduced opacities for radiation transport, thereby facilitating planet formation. We revisit the problem and establish that dust settling and grain growth also lead to outer planetary envelopes that are prone to compositional instabilities, by virtue of their inverted mean-molecular weight gradients. Under a variety of conditions,  we find that the radiative envelopes of forming planets experience compositional turbulence driven by a semi-transparent version of the thermohaline instability ('fingering convection').  
The standard double-diffusive thermohaline theory does not apply here and is replaced by a simplified first-principle treatment for the semi-transparent regime of interest.  The compositional turbulence seems efficient at mixing dust in the radiative envelopes of planets forming at super-AU distances (say $5$~AU) from a Sun-like star, but not so at sub-AU distances (say $0.2$~AU).  We also address the possibility of compositional layering in this context.  Distinct turbulent regimes for planetary envelopes growing at sub-AU vs. super-AU distances could leave an imprint on the final planets formed.   
\end{abstract}




\section{Introduction}

The core accretion scenario is the leading theoretical framework to understand the formation of gaseous giant planets. In this scenario,  a solid core accretes planetesimals and/or pebbles as it grows a massive gaseous envelope.  Eventually,  runaway gas accretion is triggered and a gas-dominated protoplanet is formed \citep{1996Icar..124...62P, 1999ApJ...521..823P, 2000ApJ...537.1013I, 2005Icar..179..415H, 2006ApJ...648..666R}.

Prior to the runaway gas accretion phase, it is possible to describe the protoplanet's structure via static or quasi-static models \citep{2014ApJ...786...21P}. The ability of the planetary envelope to cool then determines the pace of its evolution and growth, via a fundamental bottleneck set by Kelvin-Helmholz contraction  \citep{2014ApJ...786...21P, 2015ApJ...811...41L}. It depends on the magnitude of the opacity of the dust-loaded gas in the envelope, so that dust-rich envelopes, being more opaque, will cool and grow more slowly than less-opaque dust-poor envelopes.  Dust opacity also determines the critical core mass above which runaway gas accretion occurs \citep{2010ApJ...714.1343H, 2021A&A...653A.103B}.

Recently,  \cite{2014ApJ...789L..18O}  and \cite{2014A&A...572A.118M} have emphasized how the dust that is continuously accreted together with the gas from a surrounding disk will settle and experience growth in the outer protoplanet radiative envelope.  Both authors concluded that gaseous protoplanets could have envelopes that are considerably less opaque than originally thought, as a result of the strong dust opacity reduction that follows from differential settling and grain size growth.  This would  facilitate planet formation by enabling faster Kelvin-Helmholz contraction of the gaseous envelope.  It is this specific aspect of the core accretion scenario that we are revisiting in the present work.

The original core accretion scenario has been subject to many revisions and improvements over the years, in part to better address  exoplanet discoveries. These revisions include a role for planet migration \citep{2005A&A...434..343A},  pebble accretion \citep{2010A&A...520A..43O, 2012A&A...544A..32L,  2014A&A...572A.107L}, continuous envelope replenishment \citep{2015MNRAS.447.3512O, 2015ApJ...811..101F, 2018MNRAS.479..635K},  sandblasting in the convective interior \citep{2020ApJ...900...96A} and various multi-dimensional effects typically ignored in idealized 1D models \citep{2019MNRAS.487.2319B, 2019MNRAS.488.2365B}.   In this work,  we isolate the dust accretion problem that we expect to be a generic feature of the scenario even in its revised formulations.  We revisit the fate of the accreted dust, accounting for its settling and growth by following the static envelope formalism of \cite{2014ApJ...789L..18O}. Our main new finding is that the settling dust flow makes the outer  protoplanet radiative envelopes unstable to a new form of thermohaline convection, so that planetary envelopes in formation may be the site of sustained compositional turbulence.

In \S ~2, we describe our planetary envelope models.  In \S ~3 we discuss the thermohaline stability of these envelopes and the nature of the compositional turbulence that results.  We conclude by discussing several potential implications of these findings in \S ~4.

\section{Planetary Envelope Models}

\subsection{Equations}

We build planetary envelope models following closely the framework developed by \cite{2014ApJ...789L..18O} to model grain growth and settling. We use cgs units throughout our work.

The conservation of mass, momentum and energy in a self-gravitating gaseous envelope surrounding a solid core of mass $M_{\text{core}}$ are described by the set of equations
\begin{eqnarray}
\frac{d m_\text{gas}}{d r} & = & 4 \pi r^2 \rho_\text{gas} \\ 
\frac{d P}{d r} & = & - G\left[ M_{\text{core}}+ m_\text{gas}(r) \right] \frac{\rho_{\text{gas}}}{r^2} \\
\frac{d T}{d r} & = &  \frac{d P}{d r} \frac{T}{P} \nabla , 
 \end{eqnarray}
 where $\rho_{\text{gas}}$ is the gas mass density,  $m_\text{gas}(r)$ is the cumulative envelope mass up to radius $r$, $P$ is the pressure, $G$ is the gravitational constant, $T$ is the temperature. The logarithmic gradient $\nabla$ specifies the effective thermal structure of the envelope according to $\nabla = d\ln T/d \ln P = \text{min}(\nabla_{\text{rad}}, \nabla_{\text{ad}})$. For simplicity,  we adopt a fixed value $\nabla_{\text{ad}}=0.28$ for the adiabatic gradient. The radiative gradient is given by 
 \begin{equation}
    \nabla_{\text{rad}} = -\frac{3 \kappa L}{64 \pi \sigma_{\text{sb}} G M_{\text{core}}} \frac{P}{T^4}
\end{equation}
where $L$ is the core luminosity generated by accreting planetesimals, i.e.  $L = GM_{\text{core}} \dot{M}_{\text{core}}/r_{\text{core}}$, $\kappa$ is the gas+dust opacity,  and $\sigma_{\text{sb}}$ is the Stefan-Boltzmann constant. We adopt an ideal gas equation of state $P = {\rho_{\text{gas}} k_{\text{B}} T} /({\mu_{\rm gas}} m_H)$ with a gaseous mean-molecular weight $\mu_{\rm gas} =2.34$.

Following \cite{2014ApJ...789L..18O}, the gas+dust opacity $\kappa$ is given by 
\begin{equation}
    \kappa = \kappa_{\text{gas}} + \kappa_{\text{geo}}Q_e,
\end{equation}
where $\kappa_{\text{gas}} = 10^{-8}\rho_{\text{gas}}^{2/3}T^3$~cm$^2$g$^{-1}$ , and $\kappa_{\text{geo}} = 3Z_{\text{gr}}/4\rho_o$, where $\rho_o=3$g~cm$^{-3}$ is the material grain density.  The grain abundance is
\begin{equation}
    Z_{\text{gr}} = \frac{\rho_\text{gr}}{\rho_\text{gas}}
\end{equation}
and the efficiency factor $Q_e=\text{min}(0.3x,2)$ where $x=2\pi s/ \lambda_{max}$ and $\lambda_{\text{max}} = b/T$ is given by Wien's displacement law. ($b=0.2898\text{ K cm}$). 
The characteristic grain size, $s$,  used to derive the dust opacity is obtained from the grain growth model (see Eq.~\ref{eq:dust_mass} below).   

We model the settling and growth of dust grain in a static gaseous envelope according to the simplified formalism of  \cite{2014ApJ...789L..18O}.  Dust settling obeys the simple conservation equation
\begin{equation}
    \rho_\text{gr} = \frac{\dot{M}_\text{dep}}{4 \pi r^2 v_\text{settl}},
\end{equation}
where $v_\text{settl} (r)$ is the local settling velocity and $\dot{M}_\text{dep}$ is the dust mass accretion rate,  which is imposed at the outer boundary in the present models since we assume planetesimals reach the core intact in all cases (there is no dust source with the envelope itself).  The characteristic mass $m$ of the dust grain population experiencing growth obeys
\begin{equation}
    \frac{d m}{d r} = -\frac{m}{v_\text{settl} T_\text{grow}}, 
    \label{eq:dust_mass}
\end{equation}
from which one can derive the characteristic grain size,  $s=(3m/4\pi \rho_o)^{1/3}$.

For the settling velocity, we adopt $v_\text{settl} = g(r) t_\text{stop}$,  where $g(r) = GM_\text{core}/r^2$ is the local gravitational acceleration (which neglects the gaseous envelope mass). The grain stopping time, $t_\text{stop}$,  is obtained through the interpolation formula
\begin{equation}
    t_\text{stop} = \frac{\rho_o s}{c_\text{gas} \rho_\text{gas}} \cdot \text{max}\left(\frac{4}{9}, \frac{s}{l_\text{g}}\right),
\end{equation}
where $c_\text{gas} = \sqrt{k_\text{B}T/\mu_{\rm gas} m_H}$ is the sound speed, and $l_\text{g} \approx 10^{-9} \text{ cm}^{-2} \text{ g} / \rho_\text{gas} $ is the gas mean free path \citep{2020ApJ...900...96A}.  The growth rate is  $T_\text{grow}^{-1} = 3Z_\text{gr} \rho_\text{gas} \Delta v_i / \rho_o s$, with two contributions to the dust velocity differential $\Delta v_i $,  one from Brownian motion, $\Delta v_\text{bm}=\sqrt{16 k_\text{B} T/ \pi m}$ and one from differential drift, $\Delta v_\text{dd}= 0.1 v_\text{settl}$.  Like \cite{2014ApJ...789L..18O}, we simply add these two contributions to obtain $\Delta v_i$.

\subsection{Method of solutions}

We solve our system of ODEs,  made of Eqs.~1-3, together with Eq.~7 and Eq.~8, on a radial domain extending from the solid core radius at the inner boundary to the proto-planet's Hill radius at the outer boundary. For convenience, we solve Eq.~7 as an equation for the grain abundance $Z_\text{gr}$. For improved numerical convergence, we solve scaled versions of our equations and solve them in natural log radius space.  For simplicity, we fix the inner radius of our domain at  $r_{\text{in}} = 1.2\times 10^9\text{cm}$ for all our models. While this does not match the radius of the solid cores for the various masses that we explore, truncating or extending the inner radius of our solutions has little practical effect on our main conclusions,  which are largely concerned with the outermost regions of the gaseous envelopes modeled (see below).  

At the inner boundary,  we impose $m_{\text{gas}} = 0$.  We specify four outer boundary conditions, consisting of the dust mass accretion rate, $\dot{M}_\text{dep}= \dot{M}_\text{disk} = 5 \times 10^{-9} M_{\oplus} \text{ yr}^{-1}$,  the grain abundance, $Z_\text{gr} \sim 10^{-2}$-$10^{-3}$ typically,  and the gas temperature and density that match the disk values, $T_{\text{disk}} $, $\rho_{\text{disk}}$ at the Hill radius (defined below). 

The Hill and Bondi radii at semi-major axis $a$ from a Sun-like star are given by
\begin{equation}
    r_{\text{Hill}} = 2 \times 10^{11} \frac{a}{1 \text{AU}} M_{1}^{1/3}\text{ cm}
\end{equation}
\begin{equation}
    r_{\text{Bondi}} = 4 \times 10^{10} \left ( \frac{a}{1 \text{AU}} \right)^{1/2} M_{1} \text{ cm},
\end{equation}
where $M_\text{core} = M_1M_{\oplus}$.

We adopt MMSN values for the values of $T_{\text{disk}} $ and $\rho_{\text{disk}}$ imposed at the outer boundary,  so that \citep{2006ApJ...648..666R}:
\begin{equation}
    T_{\text{disk}} = 300 \left ( \frac{a}{1 \text{AU}} \right)^{-1/2} \text{ K}
\end{equation}
\begin{equation}
    \rho_{\text{disk}} = 2.4 \times 10^{-9} \left ( \frac{a}{1 \text{AU}} \right)^{-11/4} \text{ g cm}^{-3}.
\end{equation}

We solve our ODE system using the Dedalus platform \citep[v2,][]{2020PhRvR...2b3068B}, using the nonlinear boundary value problem solver.  All problem variables are transformed into their natural logarithm, except for $m_\text{gas}$.  Our implementation of approximate differentiable versions of the min and max operators is described in Appendix~A.

Unlike \cite{2014ApJ...789L..18O}, none of our models has planetesimals contributing as a source of dust within the gaseous envelope, since this feature is not essential to our main argument. In our models, planetesimals only power the core luminosity, $L$, that is conserved and transported through the gaseous envelope. Rather than fixing the core luminosity, we adopt fixed values of $\dot{M}_{\text{core}}/r_{\text{core}}$in our models, with $\dot{M}_{\text{core}} = 0.2 \times  10^{-5} M_{\oplus} \text{ yr}^{-1}$ and $r_{\text{core}}=r_{\text{in}} = 1.2\times 10^9\text{cm}$. This implies that the core accretion luminosity simply scales linearly with $M_{\text{core}}$.  We have verified that different core radii/luminosities change details of the gaseous envelope structures but do not have a strong impact on our main conclusions,  which rely on the qualitatively robust inverted mean molecular weight gradients realized in the outer envelopes, as we now describe.

\subsection{Results}

\begin{figure*}
	\includegraphics[width=\textwidth]{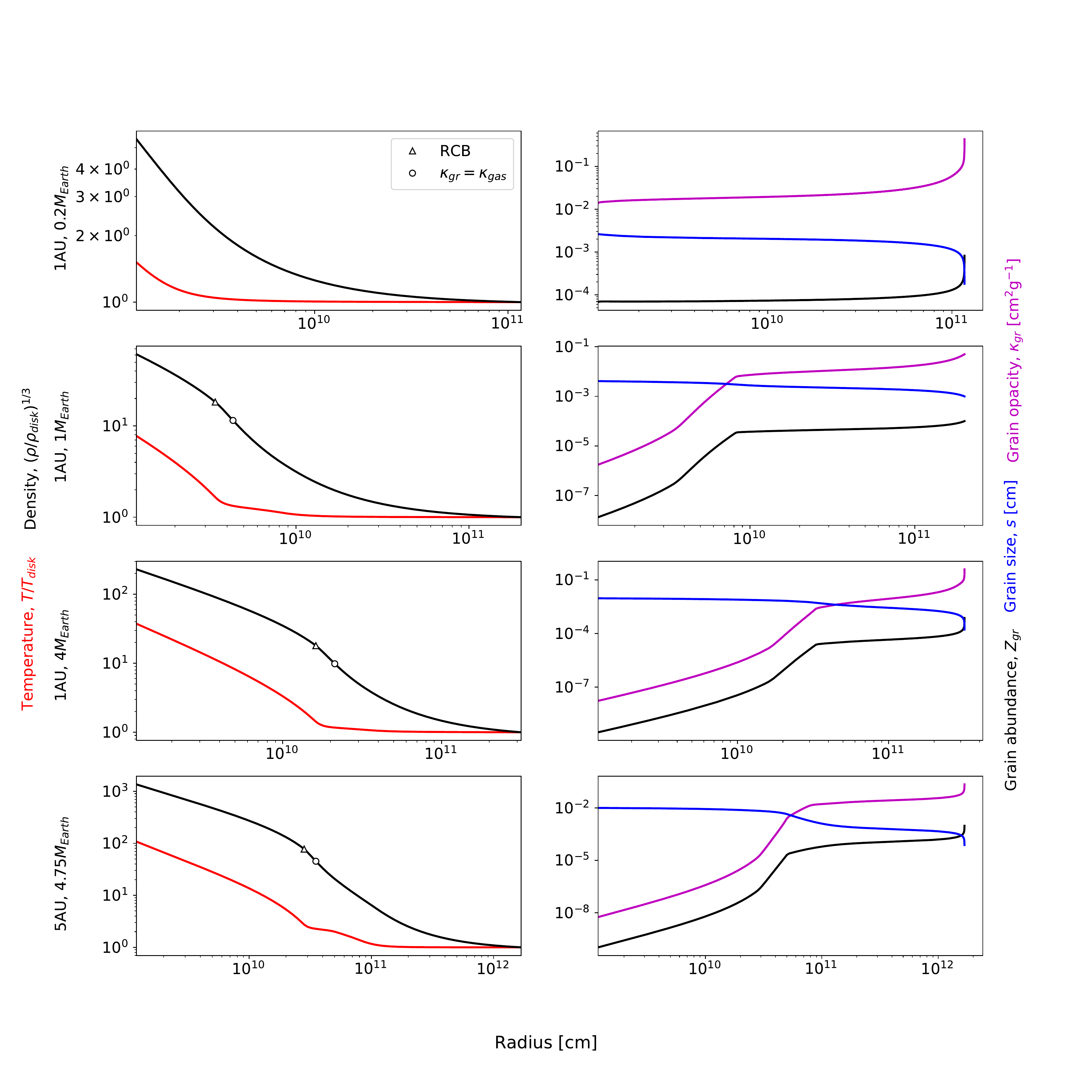}
        \caption{Temperature and density profiles (left panels) and associated settling dust properties (right panels) in four of our planetary envelope models. From top to bottom,  the panels show models for $0.2$, $1$ and $4$~M$_{\rm Earth}$ proto-planetary cores at $1$~AU,  plus a $4.75$~M$_{\rm Earth}$ core at $5$~AU from a Sun-like star.  Temperature and density profiles are scaled to the outer disk nebular values, with markers at the location of the inner convection zone (RCB) and the gas vs dust grain opacity dominance.  The grain size, abundance and resulting opacity profiles are shown in the right panels. Compare to Fig.~2 of \protect\cite{2014ApJ...789L..18O}.} 
    \label{fig:gen_profiles}
\end{figure*}

We have reproduced in detail the results of \cite{2014ApJ...789L..18O} for the various parameter values adopted in that work.  We have also surveyed a larger parameter space by building proto-planetary models over a range of semi-major axes ($0.2-5$AU), core masses ($0.2 - 5M_{\oplus} $) and boundary value for the grain abundance 
$Z_\text{gr}$ ($10^{-2} - 10^{-4} $).  Representative examples of the envelope structures and the corresponding dust properties across this parameter space are shown in Fig.~\ref{fig:gen_profiles}.

In all our models,  we find that the imposed value of $Z_\text{gr}$ quickly relaxes to a bulk value for the envelope \citep[see also][]{2014A&A...572A.118M} over a narrow dust adjustment layer close to the model outer boundary.  As mentioned by \cite{2014ApJ...789L..18O},  dust processing likely happens in the disk itself.  As a result of this edge adjustment, the specific value of $Z_\text{gr}$ imposed at the outer boundary is not particularly important to our analysis. 

By contrast,  a slowly decreasing value of $Z_\text{gr}$ with decreasing radius in the envelope's bulk is found to be present in all our envelope models, as long as the value of $Z_\text{gr}$ imposed at the outer boundary is large enough ($\sim 10^{-2}$-$10^{-4}$). This positive $Z_\text{gr}$ gradient in the bulk of the outer envelope is important since it corresponds to a mean molecular weight of the dust-loaded gas
\begin{equation}
\mu_\text{dg} = \mu_{\rm gas}  \frac{\rho_\text{gas}  + \rho_\text{gr} }{\rho_\text{gas} } = \mu_{\rm gas} (1 + Z_\text{gr})
\end{equation}  
that is increasing with radius within the envelope (see the slanted black lines in the right panels of Fig.~\ref{fig:gen_profiles}).  The resulting mean-molecular weight gradient is given by
\begin{equation}
\frac{ d  \mu_\text{dg}}{dr} = \mu_{\rm gas} \frac{d Z_\text{gr} }{dr},
\end{equation} 
which is positive.  The dust-loaded gas is thus susceptible to compositional instabilities, by virtue of the heavier fluid sitting on top of lighter fluid in the gravitational acceleration field.  We perform a detailed stability analysis of our model envelopes across the parameter space surveyed in the next section.

\section{Compositional Stability and Turbulence}

\subsection{Ledoux stability}

When typical values of $Z_\text{gr}$ are imposed at the outer boundary in our solutions (say $\sim 10^{-2}$),  we find that the outermost region of the envelope has a steep positive $\mu$-gradient that is dynamically unstable according to the Ledoux criterion (see,  e.g., the nearly vertical outer edge dust profile in the upper right panel of Fig. ~\ref{fig:gen_profiles}).  The nature of this boundary adjustment layer depends strongly on the choice of $Z_\text{gr}$ imposed at the outer boundary.  The compositional turbulence that would develop from such Ledoux-unstable conditions would presumably efficiently mix composition to more gradually match the conditions present in the disk, but we have not pursued this issue further since it also depends on the possibility of dust processing within the disk itself,  which is beyond the scope of our present work.  For the purpose of this study,  we consider this to be an edge effect that can be ignored, or adequately suppressed by lowering the outer $Z_\text{gr}$ boundary value adopted.

\subsection{Double-diffusive thermohaline stability}

Away from the edge,  in the bulk of the outer planetary envelope,  there are still 'inverted' $\mu$-gradients (i.e., positive $Z_\text{gr}$ gradients) present which are considerably milder than in the boundary adjustment layer but whose magnitude is only weakly-dependent on the specific value of $Z_\text{gr}$ imposed at the boundary.  We have verified that these milder, radially extended $\mu$-gradients are consistently weak-enough to be Ledoux stable. They can,  however,  be unstable to the double-diffusive thermohaline instability,  provided the medium offers conditions that are sufficiently doubly-diffusive (with heat diffusing much faster than compositional disparities). 

The two relevant diffusion processes are the radiative thermal diffusivity
\begin{equation}
\kappa_T \equiv \frac{16}{3} \frac{\gamma -1}{\gamma} \frac{\sigma_\text{sb} T^4}{\kappa \rho p},
\label{eq:diff_approx}
\end{equation}
here obtained in the radiative diffusion limit, and the compositional diffusivity for dust,  which presumably depends on detailed dust properties.  For concreteness, we assume in our work that the dust compositional diffusivity, $\kappa_\mu$ acts with the same efficiency as molecular viscosity and thus adopt the kinematic viscosity $\nu$ as a default value for $\kappa_\mu$.  We evaluate simply $\nu \simeq l_g c_\text{gas}$, where $l_g$ is the H$_2$ gas mean free path and $c_\text{gas}$ the sound speed \citep{2019MNRAS.485L..98M}. We return to the issue of the dust compositional diffusivity in more detail in Appendix~B,  where we describe how Brownian motion acts more slowly than viscosity in the present context and we briefly consider the possibility of dust settling within a slowly growing thermohaline finger as an additional relevant process.

Over the parameter space of interest to us here,  for protoplanetary cores with masses in the range  $0.2$-$4.75$~M$_{\rm Earth}$,  located from $0.2$ to $5$~AU from a Sun-like star,  we find typical values of the thermal diffusivity $\kappa_T$ in the range $10^{11}$-$10^{18}$~cm$^2$~s$^{-1}$ and of the kinematic viscosity $\nu$ in the range $10^2$-$10^6$~cm$^2$~s$^{-1}$. for the outer radiative envelopes. The Prandtl numbers, $Pr \equiv \nu / \kappa_T$, are in the range $10^{-7}$-$10^{-11}$. 

Throughout this work, we follow the formalism and notation of \cite{2013ApJ...768...34B} to describe the double-diffusive thermohaline instability.  In particular, the thermohaline regime of instability can be inferred from a unique number, the 'density ratio'
\begin{equation}
R_0 = \frac{\nabla - \nabla_{\rm ad}}{\nabla_{\mu}} > 0
\label{eq:Rnot}
\end{equation}   
which compares the magnitude of the stabilizing entropy gradient (numerator) to that of the destabilizing compositional gradient (denominator).  In Eq.~(\ref{eq:Rnot}),  $\nabla_\mu= d\ln \mu/d \ln P$ while the gradients $\nabla$ and $\nabla_{\rm ad}$ were defined earlier when introducing our envelope structural equations (Eqns.~1-3). Unstable thermohaline modes exist when $1< R_0 < 1/\tau$  \citep{2013ApJ...768...34B},  where the diffusivity ratio $\tau =  \kappa_\mu / \kappa_T$. Its value is set to that of the Prandtl number $Pr$ in our analysis,  since we adopt $ \kappa_\mu = \nu$ as a default (see Appendix~B for a relaxation of this assumption).

In the bulk outer regions of our proto-planetary envelopes,  we find values of $R_0$ typically ranging from $10^1$ to $10^5$,  which satisfy the condition $1< R_0 < 1/\tau (= 1/Pr)$.  This establishes that these regions are prone to double-diffusive thermohaline instabilities according to the classical double-diffusive theory.

\subsection{Semi-transparent thermohaline regime}

An additional complication in the present context is that the outer radiative envelopes of forming planets are only marginally optically thick. As a result, it is possible for the small-scale finger modes from the standard double-diffusive thermohaline theory to violate the optically-thick assumption that is implicit to the use of the radiative diffusion approximation in that theory (Eq.~\ref{eq:diff_approx}).   Indeed,  Fig. ~2 establishes that the typical scales for double-diffusive thermohaline modes are well into the optically-thin (transparent) regime,  as we now detail.

\begin{figure}
	\includegraphics[width=\columnwidth]{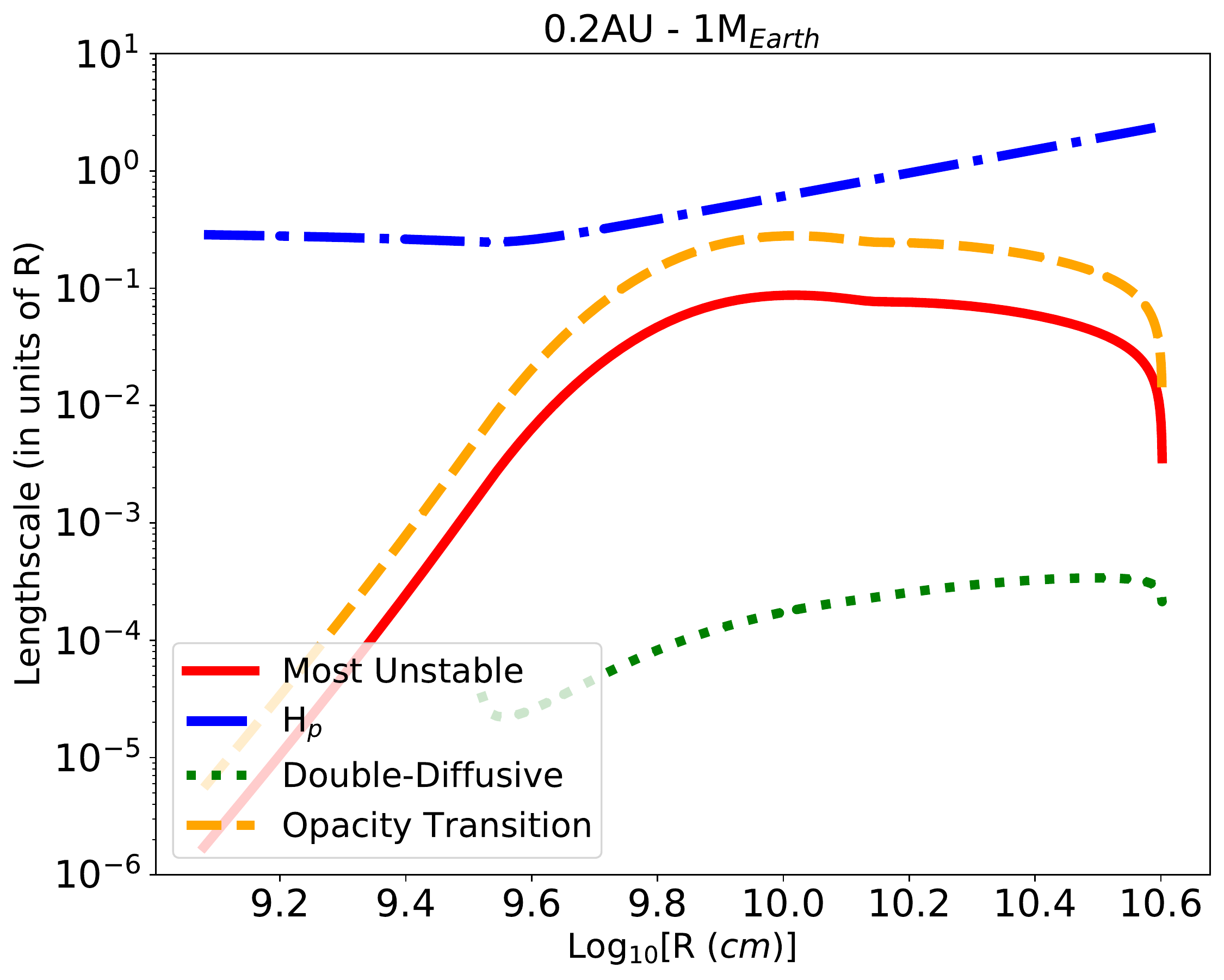}
	\includegraphics[width=\columnwidth]{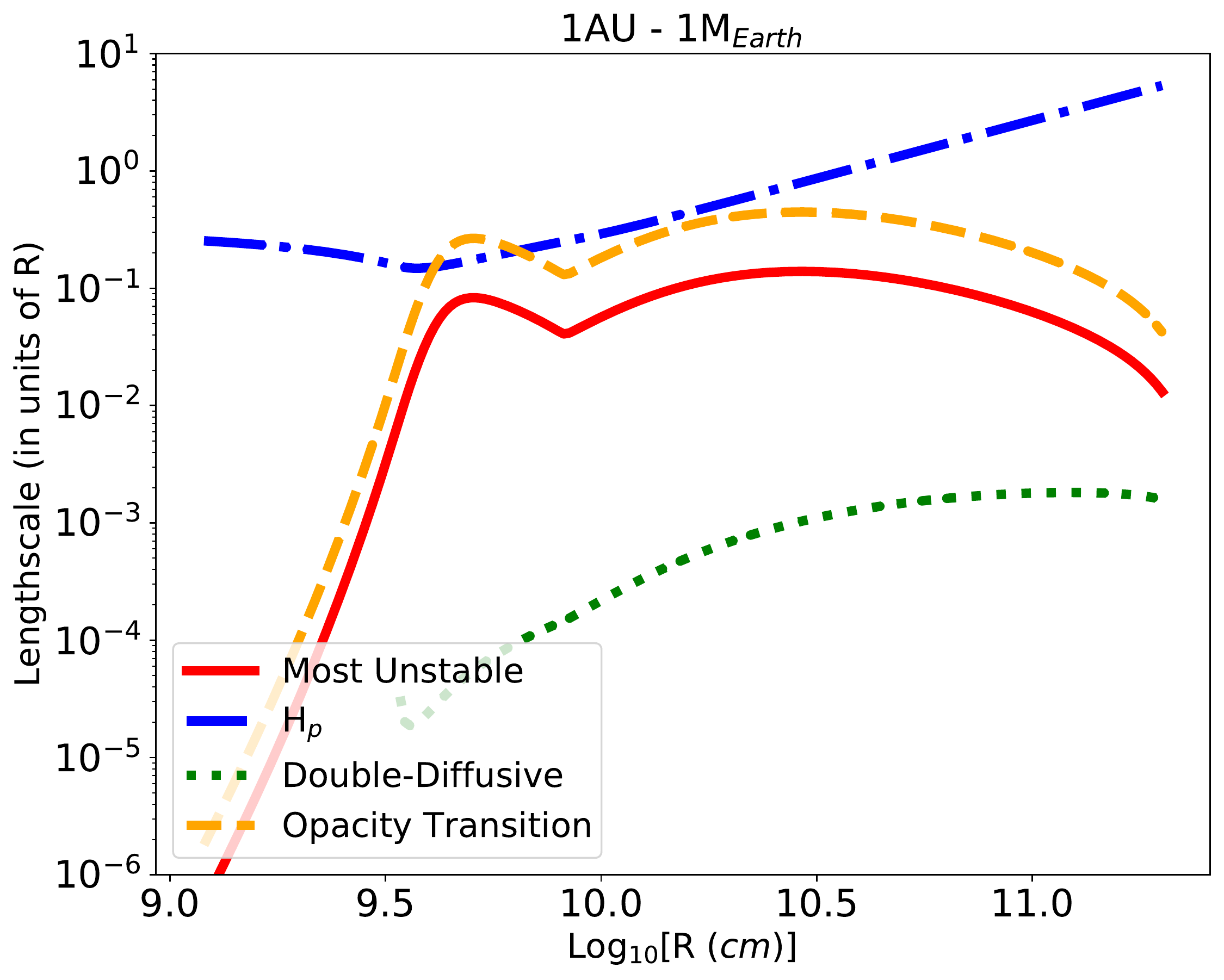}
	\includegraphics[width=\columnwidth]{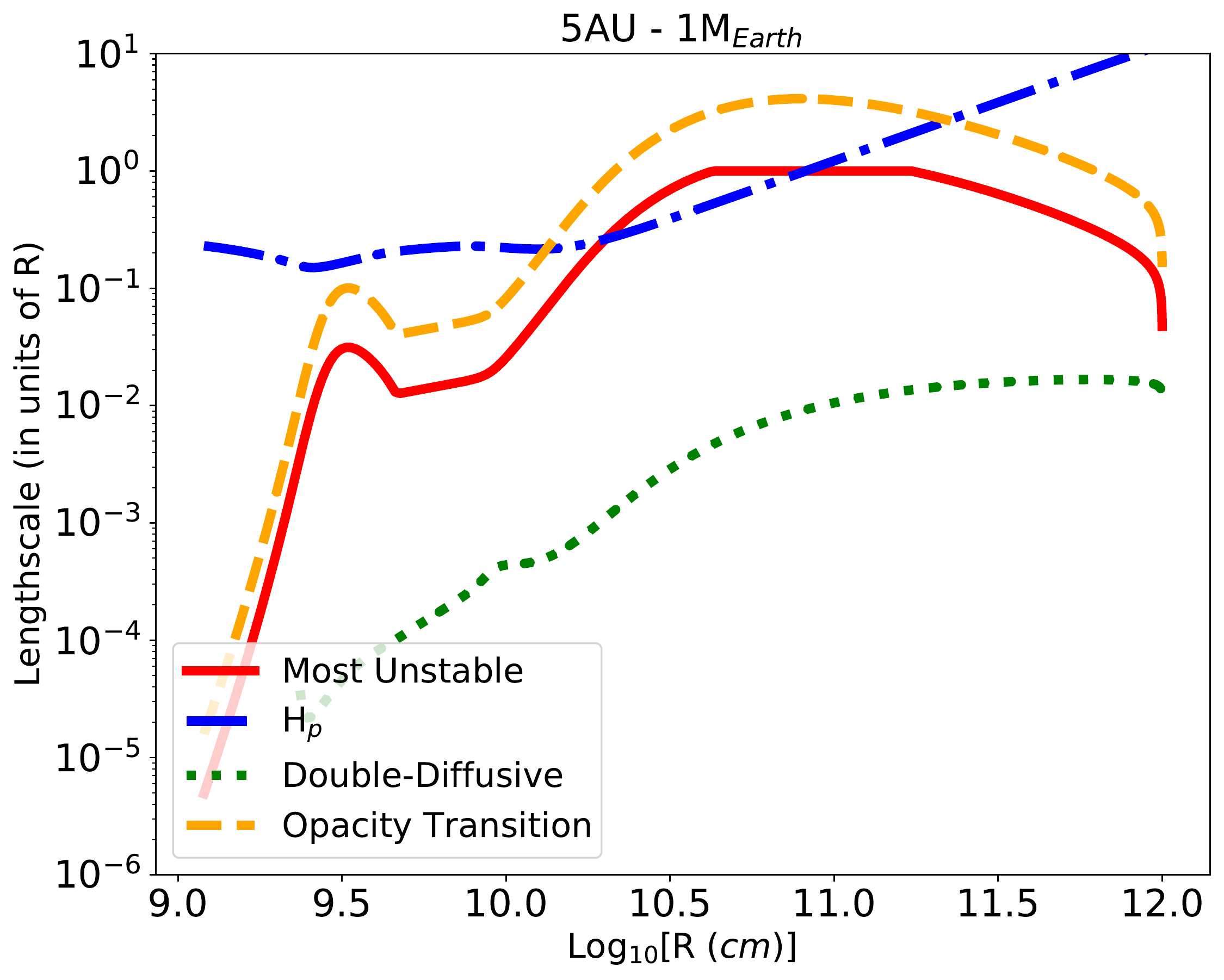}
    \caption{Lengthscales of interest in gaseous envelopes of $1$~M$_{\rm Earth}$ proto-planetary cores located  at $0.2$, $1$ and $5$~AU from a Sun-like star (top to bottom).  In each panel, the orange dashed-line shows the scale above which the dust-loaded gas becomes optically-thick.  In all cases studied, the typical finger scale for double-diffusive thermohaline modes, shown as a green dotted line,  is well into the optically-thin regime.  Our proposed gravest (most unstable) mode scale in the semi-transparent regime is shown as a red solid line. The pressure scale height, $H_p$,  is also shown as a blue dash-dotted line for reference.  Note that any quantity exceeding the local value of $H_p$ should be interpreted with caution.}
    \label{fig:lengthscale}
\end{figure}

Fig. ~2 shows profiles of various relevant length-scales for $1$~M$_{\rm Earth}$ protoplanets located at $0.2$,  $1$ and $5$~AU (top to bottom).  In each panel,  the blue dash-dotted lines show the local pressure scale height, computed as $H_p = R_\text{gas} T / g$,  as a function of logarithmic radius in the gaseous envelope above the inner solid core.  The dashed orange lines show the mean-free path of thermal photons
\begin{equation}
\lambda_{\rm mfp} = \frac{1}{\rho_\text{gas} \kappa}.
\label{eq:mfp}
\end{equation}  

The dotted green lines show the typical length-scale for double-diffusive thermohaline
fingers,  as described in \cite{2013ApJ...768...34B},
 \begin{equation}
d_\text{fing} = \left(   \frac{\nu \kappa_T}{N^2_\text{th}} \right)^{1/4},
\end{equation} 
where the (squared) thermal contribution to the Brunt-Vaissala frequency is defined as
\begin{equation}
N_\text{th}^2 = \frac{g(r)^2}{R_\text{gas} T } \left( \nabla_\text{ad} - \nabla \right).
\end{equation}

In all cases considered in this work, we find that the thermohaline finger scale is $\sim 1$-$3$ orders of magnitudes smaller than the thermal photon mean-free path,  so that thermohaline fingers are expected to be transparent with respect to the ambient thermal radiation field. This situation is reminiscent of double-diffusive shear instabilities in hot Jupiter atmospheres \citep{2021arXiv211212127M} and therefore calls for a generalization of the thermohaline theory to the semi-transparent regime.  Borrowing from our work on hot Jupiter atmospheres, we generally expect the gravest (most unstable) thermohaline mode to develop on the largest possible scale (so as to minimize the dampening effect of viscous and/or compositional microscopic diffusion) that is also able to offer as short a cooling time as possible (to help neutralize the otherwise stabilizing effect of the thermal stratification in the radiative envelope). Therefore, much like the argument developed in \cite{2021arXiv211212127M}, this leads us to posit that the largest scale offering the shortest cooling time is found right at the opacity transition scale, where the medium transitions from being transparent to being opaque. Using the definition of \cite{1957ApJ...126..202S} for the cooling time in the optically thin regime,  we have:
\begin{equation}
\tau_{\rm thin}  = \frac{C_\text{v}}{16 \kappa \sigma_\text{sb} T^3},
\end{equation} 
 where $C_\text{v}$ is the gas heat capacity at constant volume.  We infer a typical scale for the gravest mode in the semi-transparent regime
\begin{equation}
\lambda_\text{gravest} = \left(   \kappa_T \tau_{\rm thin}  \right)^{1/2},
\end{equation} 
obtained by equating the radiative diffusion time on that scale to $\tau_{\rm thin}$.   

The corresponding scale for the gravest semi-transparent thermohaline mode is shown as a solid red line in Fig.~2, capped to the maximum value of the local radius $R$. Alternatively, one could limit the gravest mode scale to the local pressure scale height,  $H_p$,  wherever $H_p < R$.  As expected from our various definitions,  the gravest mode scale is nearly equivalent to the thermal photon mean free path defined in Eq.~(\ref{eq:mfp}), shown as dashed orange lines in Fig.~2.

In \cite{2021arXiv211212127M}, several criteria were discussed for the onset of semi-transparent shear instabilities,  such as the requirement $\tau_{\rm thin} < \tau_\text{buoy} = 1 / N_\text{th}$ or the Moore-Spiegel criterion requiring  $\tau_{\rm thin} < \tau_{\rm MS}$ where
\begin{equation}
\tau_{\rm MS}  =  \frac{1}{R_0^{1/2} N_\text{th}}, 
\end{equation}
once we adapt the definition of $\tau_{\rm MS} $ to account for the destabilizing mechanism being the unstable molecular weight stratification here,  rather than the shear.  This expression for 
$\tau_{\rm MS} $ follows from replacing  the shear rate $S$ by the compositional contribution to the Brunt-Vaissala frequency, $N_{\mu}$,  and using the definition of $R_0$ in Eq.~\ref{eq:Rnot}.

Figures 3-5 show various timescales of interest for this study,  for a range of protoplanetary core masses ($0.2$-$4.75$ M$_{\rm Earth}$) located at $0.2$~AU (Fig.  3) , $1$~AU (Fig.  4) and $5$~AU (Fig.  5).  In each panel,  the red solid line shows the transparent cooling time $\tau_{\rm thin}$ as a function of logarithmic radius above the solid core boundary.  For comparison,  the buoyancy time $\tau_\text{buoy} = 1 / N_\text{th}$ and the Moore-Spiegel time $\tau_{MS}$ are also shown as dotted green and dash-dotted blue lines,  respectively.

In \cite{2021arXiv211212127M}, when discussing semi-transparent shear instabilities, the Moore-Spiegel  time was adopted as the instability threshold,  requiring that $\tau_{\rm thin} < \tau_{MS} $ for secular shear instability to occur. As we see from Figs 3- 5, following the same logic would suggest that proto-planetary envelopes at $0.2$~AU will not be subject to semi-transparent thermohaline instabilities, while the outermost regions of proto-planetary envelopes may be unstable at $1$~AU and even more so at $5$~ AU. This type of qualitative reasoning on the instability criterion was necessary for shear instabilities, because of their sub-critical nature, i.e. absent a well-defined linear stability criterion.  By contrast,  a more robust, quantitative answer can be obtained for thermohaline instabilities since they are amenable to linear stability analysis.

More specifically,  we assume that the double-diffusive linear stability analysis of \cite{2013ApJ...768...34B} for thermohaline modes remains marginally valid right at the opacity transition scale,  which is the scale at which we expect the gravest mode to grow in the semi-transparent thermohaline regime.   The reasoning behind this approach is as follows.  Perturbations on scales larger than the opacity transition scale behave as optically-thick while perturbations on smaller scales have a transparent behaviour. The effective cooling time is continuous across those scales and asymptotes to the value $\tau_{\rm thin}$ for small transparent scales. As a result,  we conjecture that there are marginally opaque scales that thermally respond in a way that is similar to the behaviour of transparent scales, so that the optically-thick linear theory remains indicative of the behaviour to be expected at the opacity transition scale.  This assumption will need to be formally evaluated in the future with dispersion relations for thermohaline modes in the transparent and semi-transparent regimes.  

Adopting this marginal validity assumption, we solve the 3rd-order dispersion relation of \cite{2013ApJ...768...34B} -- their Eq. ~19 -- for our identified gravest mode, with a length scale $\lambda_\text{gravest} $, keeping in mind that this approach pushes the limit of a linear theory that was formally derived in the double-diffusive (optically-thick) regime.  In practice,  we solve the corresponding  third-order polynomial with a standard python package ({\tt numpy.roots}) and identify the largest real root as the growth rate of our gravest mode.

The resulting growth time for the gravest mode,  $\tau_\text{fgw}$ (the inverse of the growth rate), is shown as a dashed orange line in each of the panels of Figs 3-5. In all cases, this growth time is rather slow,  as expected for a large finger with a length-scale that is well in excess of the characteristic double-diffusive scale (see Fig.~2 again). Notably,  the growth time for this gravest mode exceeds the transparent cooling time $\tau_{\rm thin}$ (red lines in Figs 3-5) by several orders of magnitude typically.  

Contrary to our initial expectations,  our dispersion relation approach also indicates that the relative scaling of $\tau_{\rm thin}$ with respect to $\tau_\text{buoy}$ and $\tau_{MS}$ has little noticeable impact on the growth time of the gravest semi-transparent mode,  in the sense that unstable semi- transparent thermohaline modes exist even when $\tau_{\rm thin} > \tau_\text{buoy}$ or  $\tau_{\rm thin} > \tau_{MS}$. 
\footnote{As a corollary, these linear stability results for thermohaline modes suggest that semi-transparent shear instability might exist beyond the qualitative instability criteria used in \cite{2021arXiv211212127M}, which would imply that turbulent vertical transport may be even more prevalent than discussed in that work.} It will be important to confirm these results in the future with dispersion relations for thermohaline modes explicitly derived for the transparent and semi-transparent regimes.

\begin{figure}
	\includegraphics[width=\columnwidth]{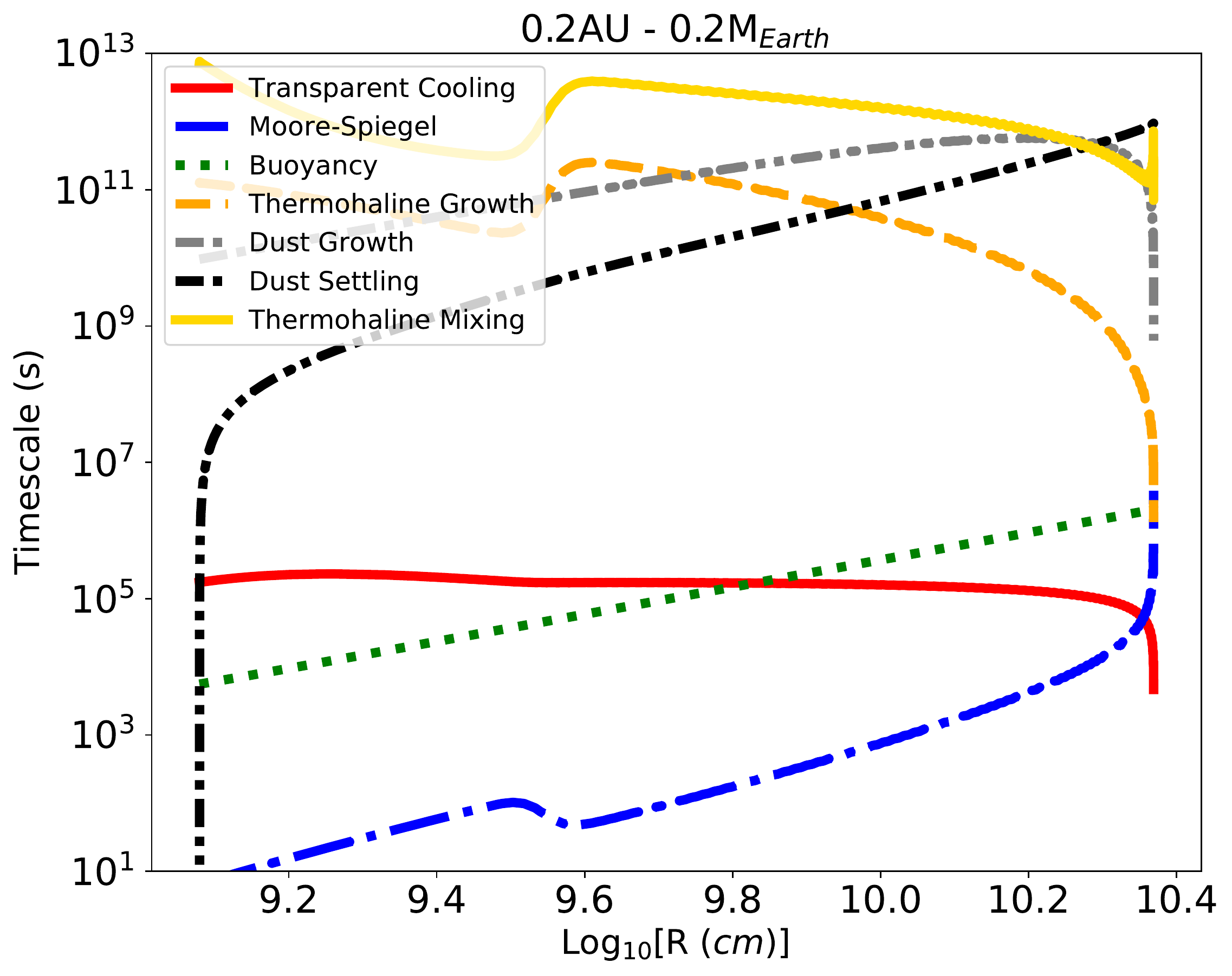}
	\includegraphics[width=\columnwidth]{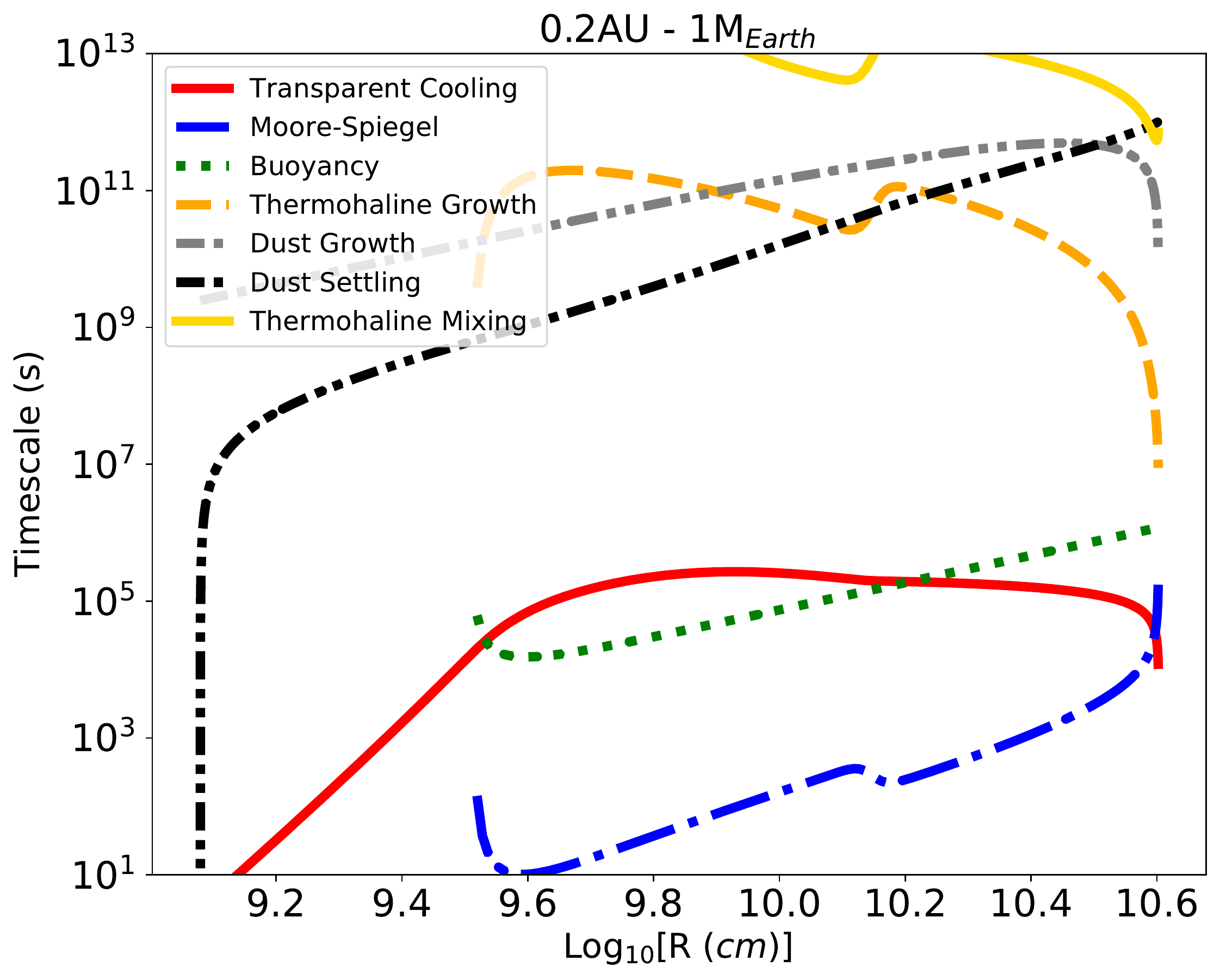}
    \caption{Timescales of interest in gaseous envelopes surrounding $0.2$ and $1$~M$_{\rm Earth}$ proto-planetary cores located at $0.2$~AU from a Sun-like star.  In each panel, the transparent cooling time,  buoyancy time and Moore-Spiegel time are shown as red solid,  green dotted and blue long-dash-dotted lines, respectively.  The semi-transparent thermohaline growth time for the gravest mode is shown as an orange dashed line. Finally,  the local thermohaline mixing time for dust (yellow solid line) is compared to the dust settling and growth times (black and grey double-dot-dashed lines, respectively).  Thermohaline activity ceases deep inside the planetary interior at the onset of regular convection and the corresponding profiles are thus truncated at the radiative-convective boundary.}
    \label{fig:time_02}
\end{figure}

\begin{figure}
	\includegraphics[width=\columnwidth]{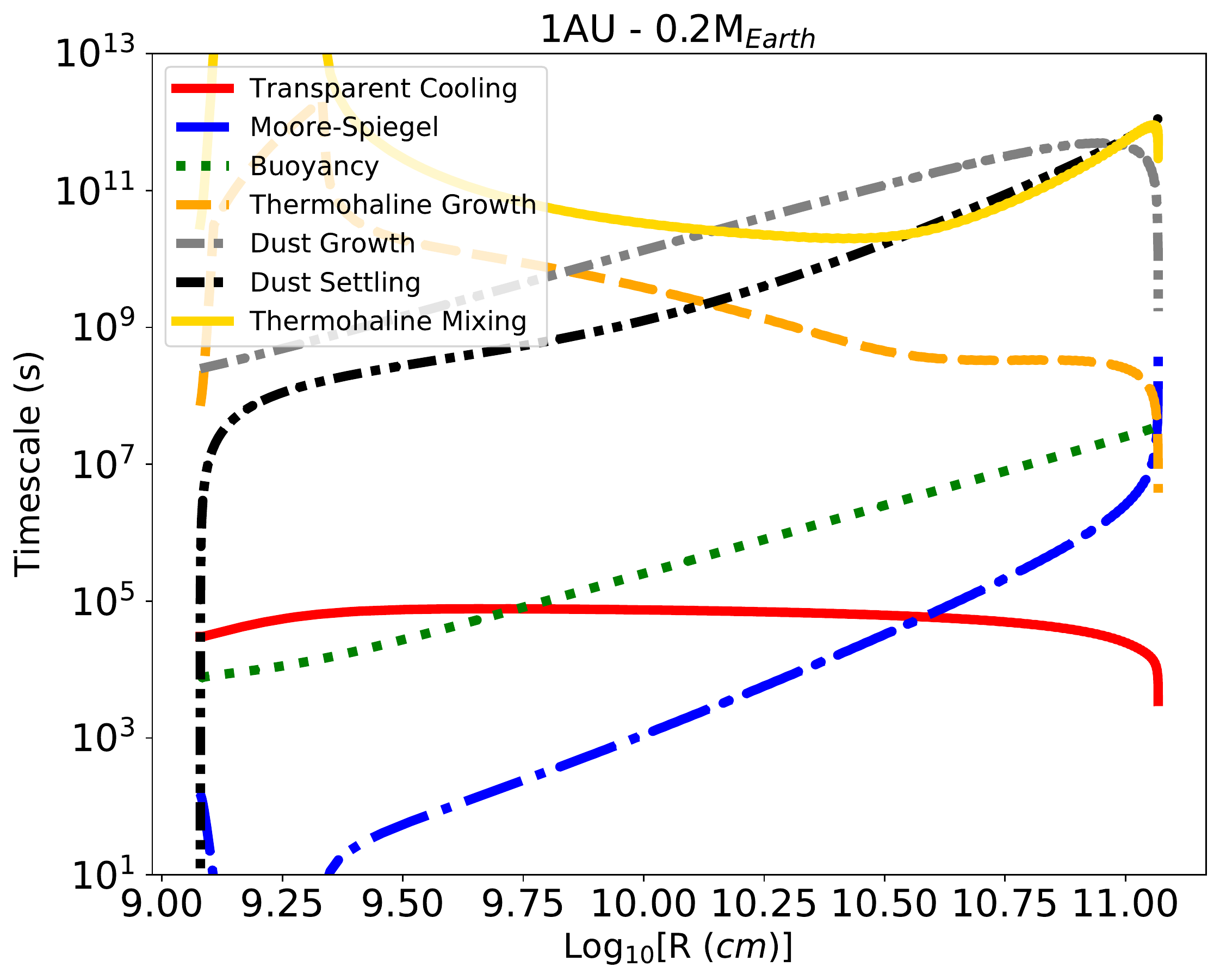}
	\includegraphics[width=\columnwidth]{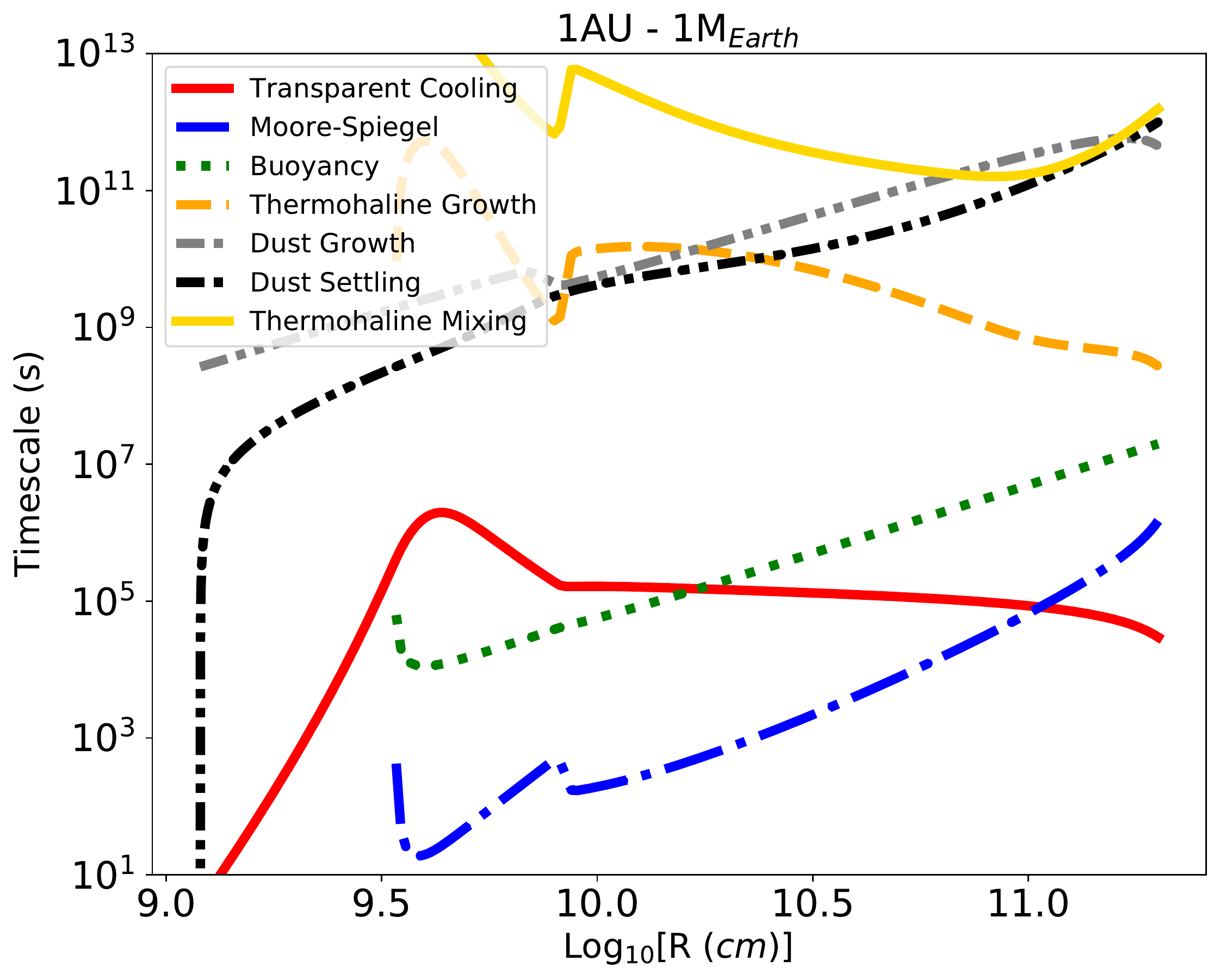}
	\includegraphics[width=\columnwidth]{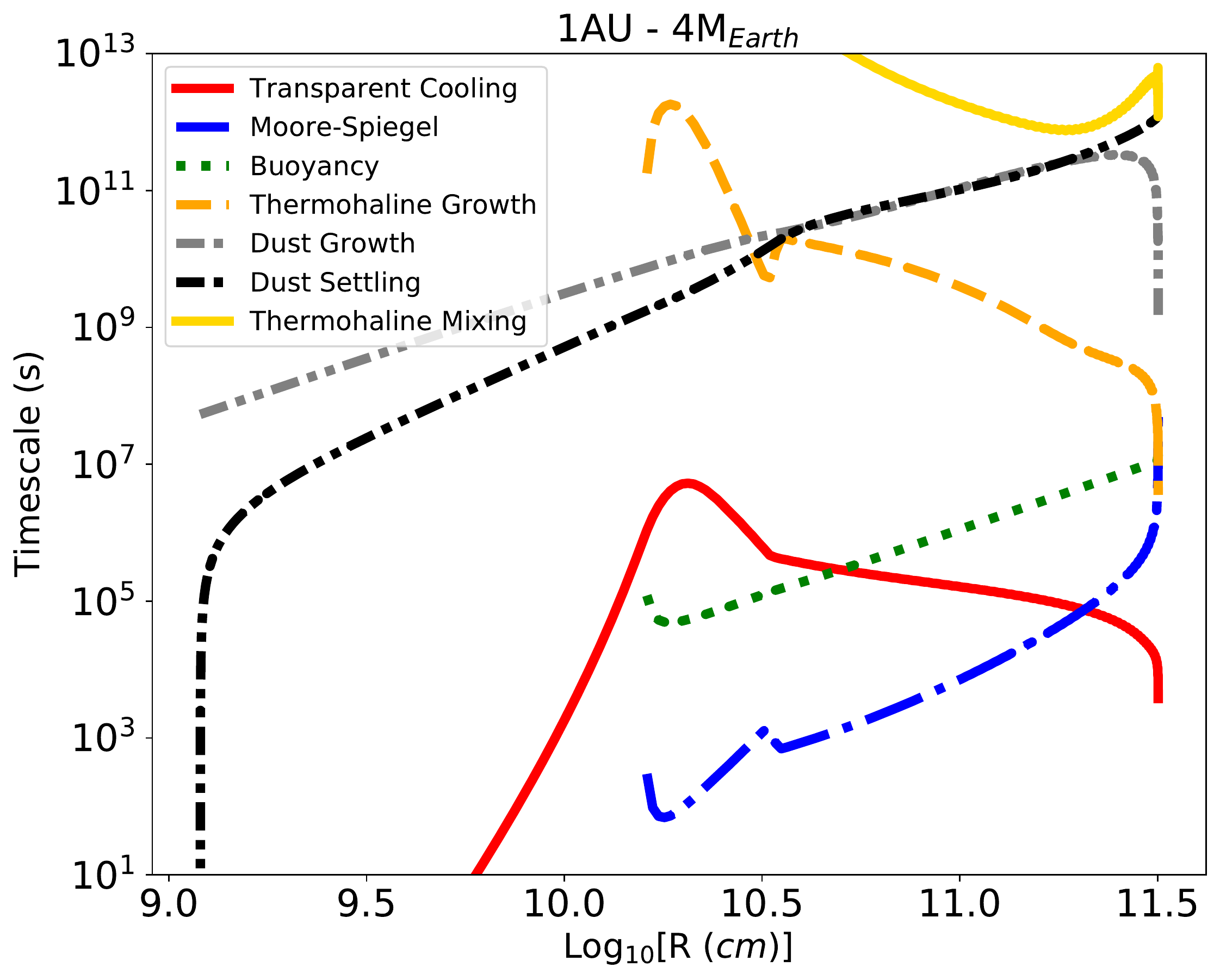}
    \caption{Timescales of interest in gaseous envelopes surrounding $0.2$, $1$ and $4$~M$_{\rm Earth}$ proto-planetary cores at $1$~AU from a Sun-like star (top to bottom). Same notation as Figure~\ref{fig:time_02}. }
    \label{fig:time_1}
\end{figure}

\begin{figure}
	\includegraphics[width=\columnwidth]{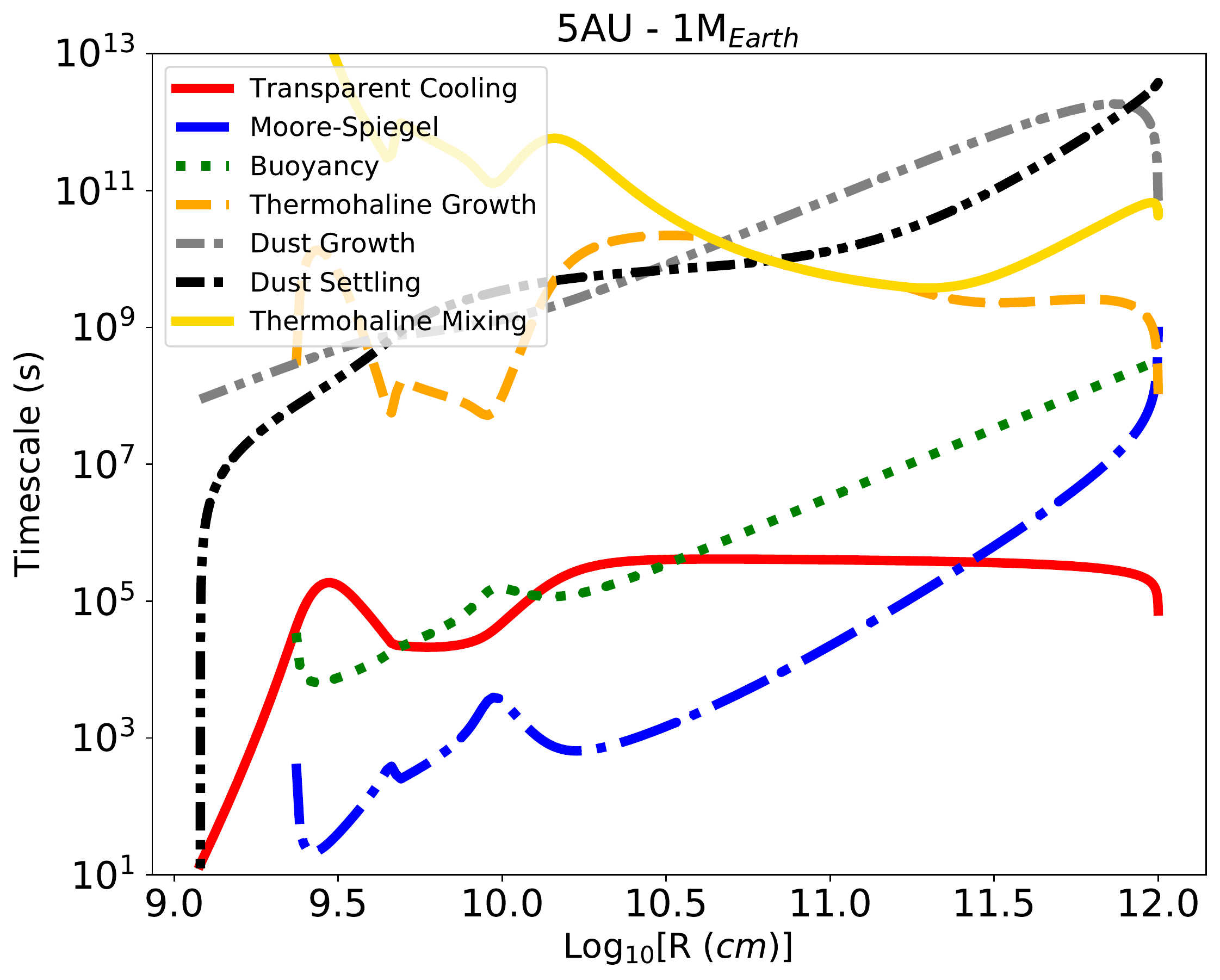}
	\includegraphics[width=\columnwidth]{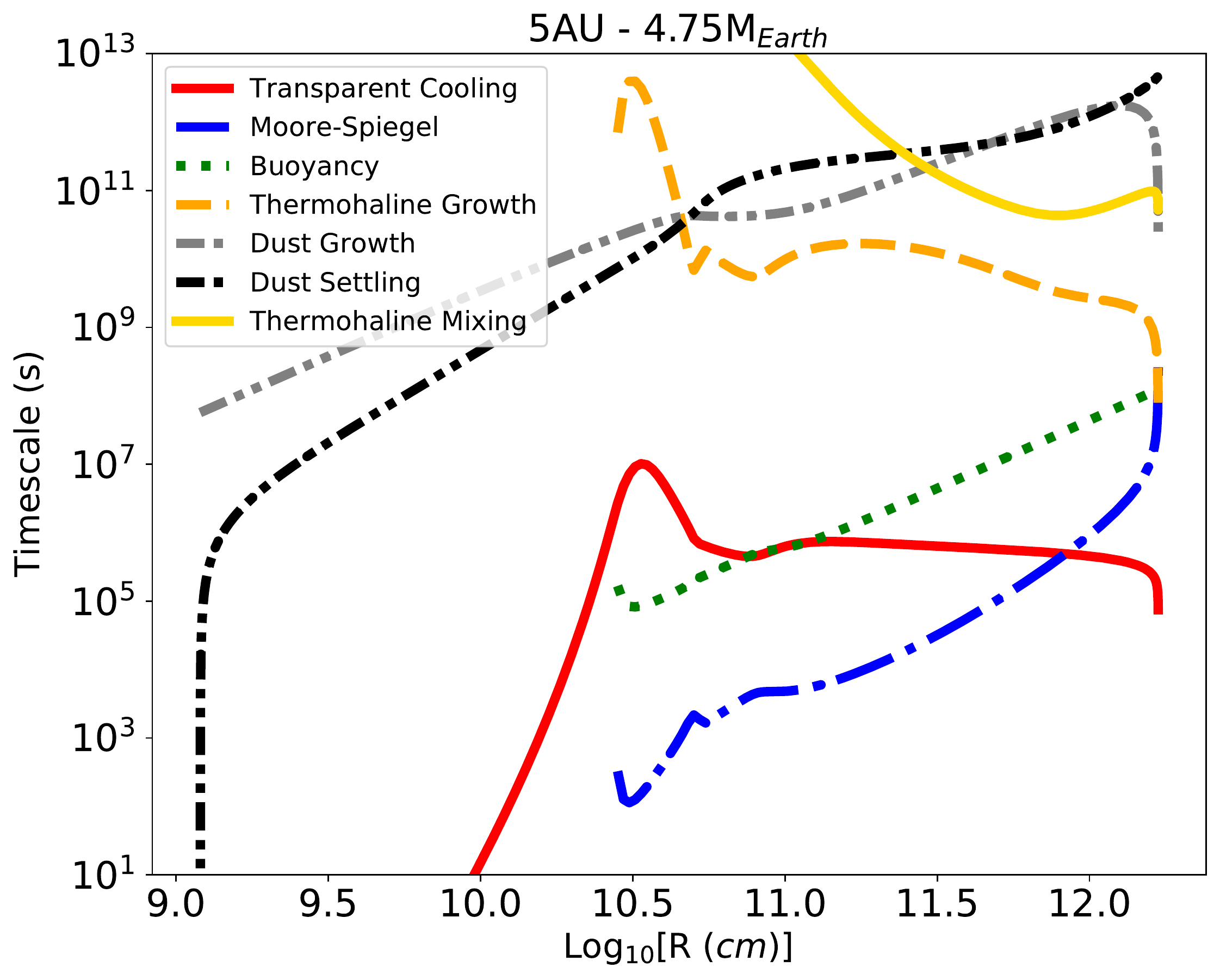}
    \caption{Timescales of interest in gaseous envelopes surrounding $1$ and $4.75$~M$_{\rm Earth}$ proto-planetary cores at $5$~AU from a Sun-like star (top to bottom). Same notation as Figure~\ref{fig:time_02}. }
    \label{fig:time_5}
\end{figure}

Equipped with estimates of the gravest mode length-scale and growth rate,  we can evaluate at linear order the magnitude of the thermohaline transport/mixing of mean-molecular weight inhomogeneities that is expected to result from this thermohaline turbulence.  For this,  we evaluate a thermohaline turbulent diffusivity as
\begin{equation}
D_\text{mix} \simeq \lambda_\text{gravest}^2 / \tau_\text{fgw} 
\end{equation}
and the corresponding local mixing time at radius $R$ in the envelope as
\begin{equation}
\tau_\text{mix} (R) \simeq R^2 / D_\text{mix}
\end{equation}

Profiles of thermohaline mixing times $\tau_\text{mix}$  are shown as solid yellow lines in each of the panels of Figs 3-5.  For comparison, the growth and settling times for dust,   $T_\text{grow}$ and $T_\text{settl} = R / v_\text{settl}$ respectively,  are also shown as grey and black dash-double-dotted lines in Figs 3-5.  In regions where $\tau_\text{mix} < T_\text{settl}$ and $< T_\text{grow}$, we expect any dust abundance gradient to be flattened (mixed) by thermohaline turbulence. \footnote{Dust settling is often faster than growth in the bulk of radiative envelopes modeled here,  except for edge effects that we neglect} As can be seen from comparing the settling vs.  mixing profiles for various protoplanetary core masses and locations in Figs 3-5,  we find that thermohaline turbulence is present in all cases modeled but the resulting mixing operates faster than dust settling (and growth) only at $5$~AU distances, with marginal competition between these two processes at $1$~AU. The mixing is comparatively inefficient at $0.2$~AU ($\tau_\text{mix} > T_\text{settl}$).

\subsection{Compositional Layering}

We can push further our use of the double-diffusive thermohaline theory for the semi-transparent regime by considering the issue of compositional layering.  It has been suggested that a compositionally turbulent medium driven by thermohaline instabilities could develop compositional layers via a collective instability when the Stern number is above unity.  Following \cite{2013ApJ...768...34B}, we compute profiles of the Stern number $A$ for our model envelopes,  using the definition for the Stern number in their Eq.~(42).

\begin{figure}
	\includegraphics[width=\columnwidth]{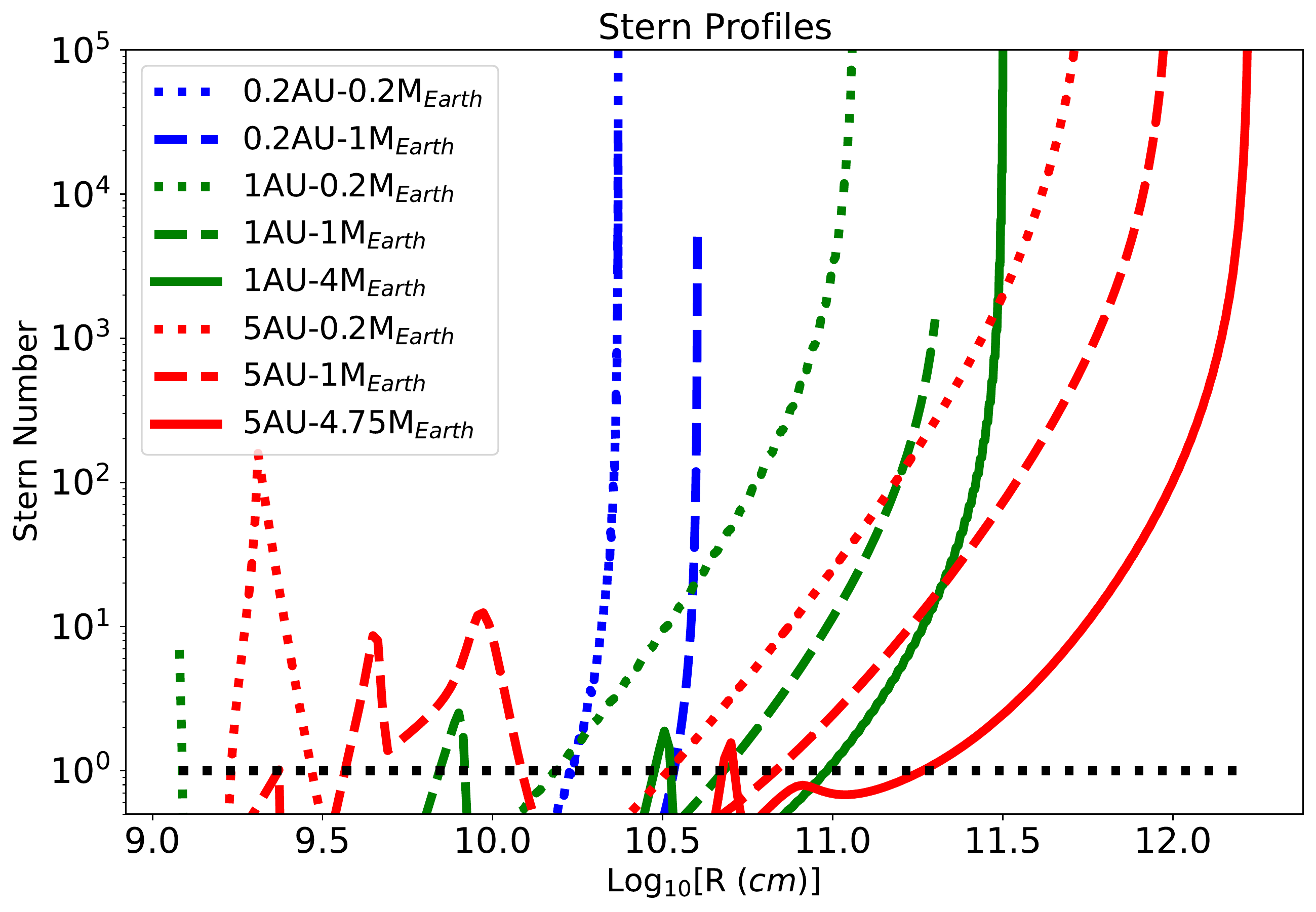}
    \caption{Stern number profiles for a variety of gaseous envelopes modelled in this work.  Excluding the steepest outer profiles, which are affected by boundary conditions,  we find that  envelopes located at super-AU distances preferentially have regions with a Stern number $A >1$,  which imply conditions prone to compositional layering.}
    \label{fig:stern}
\end{figure}

The corresponding profiles of Stern numbers are shown in Fig.~6 for our various protoplanetary envelope models,  lined up on a common logarithmic radius grid.  If we exclude nearly vertical regions of these profiles near the outer boundary, which again are caused by Ledoux-unstable edge effects, we find that the various red profiles (corresponding to various core masses located at $5$~AU) admit extended outer radiative regions with Stern number $A> 1$, while the blue profiles (for protoplanetary cores at $0.2$~AU) do not. This indicates that the conditions for compositional layering from thermohaline turbulence are preferentially met in the radiative envelopes of protoplanets forming at super-AU distances,  and perhaps marginally so at $1$~AU.

While this indicates that a different turbulent regime maybe realized at sub-AU vs super-AU distances,  one needs to remain cautious about how to interpret these results.  Our understanding of the compositional layering process remains limited \citep{2018AnRFM..50..275G}.  In the specific numerical investigation of \cite{2013ApJ...768...34B}, for example,  layering appears only for Stern numbers exceeding $100-1000$,  for reasons that remain unclear and may have numerical or physical origins, given the challenge of robustly simulating the various scales involved in the turbulent thermohaline problem.  It remains to be seen under what conditions and to what extent compositional layering can be achieved in a semi-transparent medium.

\section{Discussion and Conclusions}

We have shown that the outer radiative envelopes of gaseous planets in formation are prone to compositional instabilities, from the inverted mean-molecular weight gradients that develop in their dust-loaded gas.  Under a variety of conditions,  we find that the radiative envelopes of forming planets are thus subject to compositional turbulence driven by a semi-transparent version of the thermohaline instability ('fingering convection').  This compositional turbulence could efficiently mix dust in the envelopes of planets forming at super-AU distances from  Sun-like stars, but the mixing would not be efficient at sub-AU distances.  We also find that compositional layering resulting from a collective instability in the turbulent layers is favoured at large (super-AU) distances.  

There are various potential consequences of these results that are worth considering, even if they cannot be fully elucidated yet.  In evolutionary terms, going beyond static envelope models, it seems that the constant influx of dust-loaded gas and the dust subsequent processing in radiative envelopes will generate turbulence for planets at various stages of their formation history,  since our results hold over a large range of core masses.  

Compositional turbulence driven by double-diffusive processes is expected to be relatively mild compared to any dynamically-driven turbulence,  e.g.  turbulent convection in the interior. This might be important in determining whether the compositional turbulence is strong enough to have collisional velocities that would lead to the fragmentation of dust grains.  We can estimate typical collisional velocities in the compositional turbulence by using the properties of the gravest unstable mode: $\text{v}_\text{coll} \sim  \lambda_\text{gravest} / \tau_\text{fgw}$ is typically less than $10$~cm~s$^{-1}$. Furthermore,  dust relative velocities may be even lower than this estimate given that the dust stopping time is short relative to $\tau_\text{fgw}$, implying fairly coherent dust motion within the turbulent fluid. This suggests that systematic collisional grinding is not expected in the compositional turbulence of interest here \citep[unlike the convective interior,][]{2020ApJ...900...96A}.  

Assuming no collisional fragmentation occurs, the main effect of compositional turbulence would then be its tendency to mix the larger (processed) dust grains found at depth with the smaller (unprocessed) dust grains found further out in the envelope.  We expect mixing in the thermohaline-unstable region to combine with additional (stronger) mixing in the Ledoux-unstable border region. \footnote{We note that collisional fragmentation may be more likely to happen in Ledoux-unstable regions.} This composite mixing by compositional turbulence will act to eliminate the original instabilities driving the turbulence and should thus flatten the grain abundance profile in the outer planetary envelope.  This would tend to also flatten the closely related opacity profile of the outer envelopes (see Fig.~1, black and pink solid lines) and, if strong enough,  could potentially counter the opacity reduction effect emphasized by \cite{2014ApJ...789L..18O} and \cite{2014A&A...572A.118M}.   A quantitative evaluation of these effects will likely require solutions to an advection-diffusion equation for the dust-grain population.   We note,  as a further consequence,  that more opaque envelopes would build up mass at slower rates.  Our results thus suggest the possibility of comparatively faster planet formation at sub-AU distances than at super-AU distances, although it is clear that other effects not accounted for in our work can greatly impact the ability of any forming planet to grow its gaseous envelope \citep[e.g.,][]{2015MNRAS.447.3512O}.    

The consequences of compositional layering,  if it happens,  are somewhat unclear. It would presumably impact envelope cooling by reducing heat transport,  which would further slow down the gaseous envelope buildup.  One also wonders whether the compositional layers could survive the later stages of planet formation (e.g. the fast runaway gas accretion stage) and become permanent features of the final planetary envelopes thus formed.  Compositional layers have been proposed to better explain the structure and evolution of Solar System and extrasolar gaseous planets \citep{2012A&A...540A..20L, 2013NatGe...6..347L} but their origin remains unclear \citep[see also][]{2022arXiv220412643F}.

Finally,  we note that forming planets are expected to rotate fast from the buildup of angular momentum that comes with the gas being accreted \citep{2019MNRAS.487.2319B, 2019MNRAS.488.2365B}.  Interestingly,  \cite{2018ApJ...862..136S} have studied the role of rotation on thermohaline modes and compositional turbulence and have pointed out that rotation may strengthen thermohaline instabilities by limiting the transverse shear instabilities that are thought to saturate the growth of thermohaline fingers.  

These various points suggest that there are several interesting avenues of exploration going beyond our first-level analysis that will be important to develop a better understanding of the role of compositional turbulence in the gaseous envelopes of forming planets.


\section*{Data Availability Statement}
The data underlying this article will be shared on reasonable request to the corresponding author (KM).

\section*{Acknowledgements}
We thank the anonymous referee for constructive comments that helped improve the manuscript. KM thanks Jeremy Goodman for advice on the literature regarding dust-gas coupling. KM is supported by the National Science and Engineering Research Council of Canada. This work has made extensive use of the following software packages: {\tt numpy}, {\tt matplotlib}, {\tt Dedalus}.

We would furthermore like to acknowledge that this work was performed on land which for thousands of years has been the traditional land of the Huron-Wendat, the Seneca and the Mississaugas of the Credit. Today this meeting place is still the home to many indigenous people from across Turtle Island and we are grateful to have the opportunity to work on this land.




\bibliographystyle{mnras}
\bibliography{dusty_planet_form_v2}

\begin{thebibliography}{}
\makeatletter
\relax
\def\mn@urlcharsother{\let\do\@makeother \do\$\do\&\do\#\do\^\do\_\do\%\do\~}
\def\mn@doi{\begingroup\mn@urlcharsother \@ifnextchar [ {\mn@doi@}
  {\mn@doi@[]}}
\def\mn@doi@[#1]#2{\def\@tempa{#1}\ifx\@tempa\@empty \href
  {http://dx.doi.org/#2} {doi:#2}\else \href {http://dx.doi.org/#2} {#1}\fi
  \endgroup}
\def\mn@eprint#1#2{\mn@eprint@#1:#2::\@nil}
\def\mn@eprint@arXiv#1{\href {http://arxiv.org/abs/#1} {{\tt arXiv:#1}}}
\def\mn@eprint@dblp#1{\href {http://dblp.uni-trier.de/rec/bibtex/#1.xml}
  {dblp:#1}}
\def\mn@eprint@#1:#2:#3:#4\@nil{\def\@tempa {#1}\def\@tempb {#2}\def\@tempc
  {#3}\ifx \@tempc \@empty \let \@tempc \@tempb \let \@tempb \@tempa \fi \ifx
  \@tempb \@empty \def\@tempb {arXiv}\fi \@ifundefined
  {mn@eprint@\@tempb}{\@tempb:\@tempc}{\expandafter \expandafter \csname
  mn@eprint@\@tempb\endcsname \expandafter{\@tempc}}}

\bibitem[\protect\citeauthoryear{{Ali-Dib} \& {Thompson}}{{Ali-Dib} \&
  {Thompson}}{2020}]{2020ApJ...900...96A}
{Ali-Dib} M.,  {Thompson} C.,  2020, \mn@doi [\apj] {10.3847/1538-4357/aba521},
  \href {https://ui.adsabs.harvard.edu/abs/2020ApJ...900...96A} {900, 96}

\bibitem[\protect\citeauthoryear{{Alibert}, {Mordasini}, {Benz}  \&
  {Winisdoerffer}}{{Alibert} et~al.}{2005}]{2005A&A...434..343A}
{Alibert} Y.,  {Mordasini} C.,  {Benz} W.,   {Winisdoerffer} C.,  2005, \mn@doi
  [\aap] {10.1051/0004-6361:20042032}, \href
  {https://ui.adsabs.harvard.edu/abs/2005A&A...434..343A} {434, 343}

\bibitem[\protect\citeauthoryear{{Armitage}}{{Armitage}}{2007}]{2007astro.ph..1485A}
{Armitage} P.~J.,  2007, arXiv e-prints, \href
  {https://ui.adsabs.harvard.edu/abs/2007astro.ph..1485A} {pp
  astro--ph/0701485}

\bibitem[\protect\citeauthoryear{{B{\'e}thune} \& {Rafikov}}{{B{\'e}thune} \&
  {Rafikov}}{2019a}]{2019MNRAS.487.2319B}
{B{\'e}thune} W.,  {Rafikov} R.~R.,  2019a, \mn@doi [\mnras]
  {10.1093/mnras/stz1427}, \href
  {https://ui.adsabs.harvard.edu/abs/2019MNRAS.487.2319B} {487, 2319}

\bibitem[\protect\citeauthoryear{{B{\'e}thune} \& {Rafikov}}{{B{\'e}thune} \&
  {Rafikov}}{2019b}]{2019MNRAS.488.2365B}
{B{\'e}thune} W.,  {Rafikov} R.~R.,  2019b, \mn@doi [\mnras]
  {10.1093/mnras/stz1870}, \href
  {https://ui.adsabs.harvard.edu/abs/2019MNRAS.488.2365B} {488, 2365}

\bibitem[\protect\citeauthoryear{{Blum}, {Wurm}, {Kempf}  \& {Henning}}{{Blum}
  et~al.}{1996}]{1996Icar..124..441B}
{Blum} J.,  {Wurm} G.,  {Kempf} S.,   {Henning} T.,  1996, \mn@doi [\icarus]
  {10.1006/icar.1996.0221}, \href
  {https://ui.adsabs.harvard.edu/abs/1996Icar..124..441B} {124, 441}

\bibitem[\protect\citeauthoryear{{Brouwers}, {Ormel}, {Bonsor}  \&
  {Vazan}}{{Brouwers} et~al.}{2021}]{2021A&A...653A.103B}
{Brouwers} M.~G.,  {Ormel} C.~W.,  {Bonsor} A.,   {Vazan} A.,  2021, \mn@doi
  [\aap] {10.1051/0004-6361/202140476}, \href
  {https://ui.adsabs.harvard.edu/abs/2021A&A...653A.103B} {653, A103}

\bibitem[\protect\citeauthoryear{{Brown}, {Garaud}  \& {Stellmach}}{{Brown}
  et~al.}{2013}]{2013ApJ...768...34B}
{Brown} J.~M.,  {Garaud} P.,   {Stellmach} S.,  2013, \mn@doi [\apj]
  {10.1088/0004-637X/768/1/34}, \href
  {https://ui.adsabs.harvard.edu/abs/2013ApJ...768...34B} {768, 34}

\bibitem[\protect\citeauthoryear{{Burns}, {Vasil}, {Oishi}, {Lecoanet}  \&
  {Brown}}{{Burns} et~al.}{2020}]{2020PhRvR...2b3068B}
{Burns} K.~J.,  {Vasil} G.~M.,  {Oishi} J.~S.,  {Lecoanet} D.,   {Brown} B.~P.,
   2020, \mn@doi [Physical Review Research] {10.1103/PhysRevResearch.2.023068},
  \href {https://ui.adsabs.harvard.edu/abs/2020PhRvR...2b3068B} {2, 023068}

\bibitem[\protect\citeauthoryear{{Fuentes}, {Cumming}  \& {Anders}}{{Fuentes}
  et~al.}{2022}]{2022arXiv220412643F}
{Fuentes} J.~R.,  {Cumming} A.,   {Anders} E.~H.,  2022, arXiv e-prints, \href
  {https://ui.adsabs.harvard.edu/abs/2022arXiv220412643F} {p. arXiv:2204.12643}

\bibitem[\protect\citeauthoryear{{Fung}, {Artymowicz}  \& {Wu}}{{Fung}
  et~al.}{2015}]{2015ApJ...811..101F}
{Fung} J.,  {Artymowicz} P.,   {Wu} Y.,  2015, \mn@doi [\apj]
  {10.1088/0004-637X/811/2/101}, \href
  {https://ui.adsabs.harvard.edu/abs/2015ApJ...811..101F} {811, 101}

\bibitem[\protect\citeauthoryear{{Garaud}}{{Garaud}}{2018}]{2018AnRFM..50..275G}
{Garaud} P.,  2018, \mn@doi [Annual Review of Fluid Mechanics]
  {10.1146/annurev-fluid-122316-045234}, \href
  {https://ui.adsabs.harvard.edu/abs/2018AnRFM..50..275G} {50, 275}

\bibitem[\protect\citeauthoryear{{Hori} \& {Ikoma}}{{Hori} \&
  {Ikoma}}{2010}]{2010ApJ...714.1343H}
{Hori} Y.,  {Ikoma} M.,  2010, \mn@doi [\apj] {10.1088/0004-637X/714/2/1343},
  \href {https://ui.adsabs.harvard.edu/abs/2010ApJ...714.1343H} {714, 1343}

\bibitem[\protect\citeauthoryear{{Hubickyj}, {Bodenheimer}  \&
  {Lissauer}}{{Hubickyj} et~al.}{2005}]{2005Icar..179..415H}
{Hubickyj} O.,  {Bodenheimer} P.,   {Lissauer} J.~J.,  2005, \mn@doi [\icarus]
  {10.1016/j.icarus.2005.06.021}, \href
  {https://ui.adsabs.harvard.edu/abs/2005Icar..179..415H} {179, 415}

\bibitem[\protect\citeauthoryear{{Ikoma}, {Nakazawa}  \& {Emori}}{{Ikoma}
  et~al.}{2000}]{2000ApJ...537.1013I}
{Ikoma} M.,  {Nakazawa} K.,   {Emori} H.,  2000, \mn@doi [\apj]
  {10.1086/309050}, \href
  {https://ui.adsabs.harvard.edu/abs/2000ApJ...537.1013I} {537, 1013}

\bibitem[\protect\citeauthoryear{{Kurokawa} \& {Tanigawa}}{{Kurokawa} \&
  {Tanigawa}}{2018}]{2018MNRAS.479..635K}
{Kurokawa} H.,  {Tanigawa} T.,  2018, \mn@doi [\mnras] {10.1093/mnras/sty1498},
  \href {https://ui.adsabs.harvard.edu/abs/2018MNRAS.479..635K} {479, 635}

\bibitem[\protect\citeauthoryear{{Lambrechts} \& {Johansen}}{{Lambrechts} \&
  {Johansen}}{2012}]{2012A&A...544A..32L}
{Lambrechts} M.,  {Johansen} A.,  2012, \mn@doi [\aap]
  {10.1051/0004-6361/201219127}, \href
  {https://ui.adsabs.harvard.edu/abs/2012A&A...544A..32L} {544, A32}

\bibitem[\protect\citeauthoryear{{Lambrechts} \& {Johansen}}{{Lambrechts} \&
  {Johansen}}{2014}]{2014A&A...572A.107L}
{Lambrechts} M.,  {Johansen} A.,  2014, \mn@doi [\aap]
  {10.1051/0004-6361/201424343}, \href
  {https://ui.adsabs.harvard.edu/abs/2014A&A...572A.107L} {572, A107}

\bibitem[\protect\citeauthoryear{{Leconte} \& {Chabrier}}{{Leconte} \&
  {Chabrier}}{2012}]{2012A&A...540A..20L}
{Leconte} J.,  {Chabrier} G.,  2012, \mn@doi [\aap]
  {10.1051/0004-6361/201117595}, \href
  {https://ui.adsabs.harvard.edu/abs/2012A&A...540A..20L} {540, A20}

\bibitem[\protect\citeauthoryear{{Leconte} \& {Chabrier}}{{Leconte} \&
  {Chabrier}}{2013}]{2013NatGe...6..347L}
{Leconte} J.,  {Chabrier} G.,  2013, \mn@doi [Nature Geoscience]
  {10.1038/ngeo1791}, \href
  {https://ui.adsabs.harvard.edu/abs/2013NatGe...6..347L} {6, 347}

\bibitem[\protect\citeauthoryear{{Lee} \& {Chiang}}{{Lee} \&
  {Chiang}}{2015}]{2015ApJ...811...41L}
{Lee} E.~J.,  {Chiang} E.,  2015, \mn@doi [\apj] {10.1088/0004-637X/811/1/41},
  \href {https://ui.adsabs.harvard.edu/abs/2015ApJ...811...41L} {811, 41}

\bibitem[\protect\citeauthoryear{{Menou}}{{Menou}}{2019}]{2019MNRAS.485L..98M}
{Menou} K.,  2019, \mn@doi [\mnras] {10.1093/mnrasl/slz041}, \href
  {https://ui.adsabs.harvard.edu/abs/2019MNRAS.485L..98M} {485, L98}

\bibitem[\protect\citeauthoryear{{Menou}}{{Menou}}{2021}]{2021arXiv211212127M}
{Menou} K.,  2021, arXiv e-prints, \href
  {https://ui.adsabs.harvard.edu/abs/2021arXiv211212127M} {p. arXiv:2112.12127}

\bibitem[\protect\citeauthoryear{{Mordasini}}{{Mordasini}}{2014}]{2014A&A...572A.118M}
{Mordasini} C.,  2014, \mn@doi [\aap] {10.1051/0004-6361/201423702}, \href
  {https://ui.adsabs.harvard.edu/abs/2014A&A...572A.118M} {572, A118}

\bibitem[\protect\citeauthoryear{{Ormel}}{{Ormel}}{2014}]{2014ApJ...789L..18O}
{Ormel} C.~W.,  2014, \mn@doi [\apjl] {10.1088/2041-8205/789/1/L18}, \href
  {https://ui.adsabs.harvard.edu/abs/2014ApJ...789L..18O} {789, L18}

\bibitem[\protect\citeauthoryear{{Ormel} \& {Klahr}}{{Ormel} \&
  {Klahr}}{2010}]{2010A&A...520A..43O}
{Ormel} C.~W.,  {Klahr} H.~H.,  2010, \mn@doi [\aap]
  {10.1051/0004-6361/201014903}, \href
  {https://ui.adsabs.harvard.edu/abs/2010A&A...520A..43O} {520, A43}

\bibitem[\protect\citeauthoryear{{Ormel}, {Shi}  \& {Kuiper}}{{Ormel}
  et~al.}{2015}]{2015MNRAS.447.3512O}
{Ormel} C.~W.,  {Shi} J.-M.,   {Kuiper} R.,  2015, \mn@doi [\mnras]
  {10.1093/mnras/stu2704}, \href
  {https://ui.adsabs.harvard.edu/abs/2015MNRAS.447.3512O} {447, 3512}

\bibitem[\protect\citeauthoryear{{Papaloizou} \& {Terquem}}{{Papaloizou} \&
  {Terquem}}{1999}]{1999ApJ...521..823P}
{Papaloizou} J. C.~B.,  {Terquem} C.,  1999, \mn@doi [\apj] {10.1086/307581},
  \href {https://ui.adsabs.harvard.edu/abs/1999ApJ...521..823P} {521, 823}

\bibitem[\protect\citeauthoryear{{Piso} \& {Youdin}}{{Piso} \&
  {Youdin}}{2014}]{2014ApJ...786...21P}
{Piso} A.-M.~A.,  {Youdin} A.~N.,  2014, \mn@doi [\apj]
  {10.1088/0004-637X/786/1/21}, \href
  {https://ui.adsabs.harvard.edu/abs/2014ApJ...786...21P} {786, 21}

\bibitem[\protect\citeauthoryear{{Pollack}, {Hubickyj}, {Bodenheimer},
  {Lissauer}, {Podolak}  \& {Greenzweig}}{{Pollack}
  et~al.}{1996}]{1996Icar..124...62P}
{Pollack} J.~B.,  {Hubickyj} O.,  {Bodenheimer} P.,  {Lissauer} J.~J.,
  {Podolak} M.,   {Greenzweig} Y.,  1996, \mn@doi [\icarus]
  {10.1006/icar.1996.0190}, \href
  {https://ui.adsabs.harvard.edu/abs/1996Icar..124...62P} {124, 62}

\bibitem[\protect\citeauthoryear{{Rafikov}}{{Rafikov}}{2006}]{2006ApJ...648..666R}
{Rafikov} R.~R.,  2006, \mn@doi [\apj] {10.1086/505695}, \href
  {https://ui.adsabs.harvard.edu/abs/2006ApJ...648..666R} {648, 666}

\bibitem[\protect\citeauthoryear{{Sengupta} \& {Garaud}}{{Sengupta} \&
  {Garaud}}{2018}]{2018ApJ...862..136S}
{Sengupta} S.,  {Garaud} P.,  2018, \mn@doi [\apj] {10.3847/1538-4357/aacbc8},
  \href {https://ui.adsabs.harvard.edu/abs/2018ApJ...862..136S} {862, 136}

\bibitem[\protect\citeauthoryear{{Spiegel}}{{Spiegel}}{1957}]{1957ApJ...126..202S}
{Spiegel} E.~A.,  1957, \mn@doi [\apj] {10.1086/146386}, \href
  {https://ui.adsabs.harvard.edu/abs/1957ApJ...126..202S} {126, 202}

\makeatother
\end{thebibliography}


\appendix

\section{Continuous Min/Max functions}

The max and min functions used in some of our expressions are not differentiable, so we use the smooth approximations: 
\begin{equation}
    \text{max}(x,y) = \frac{1}{2}\left( x^a + y^a\right)^{1/a}
\end{equation}
\begin{equation}
    \text{min}(x,y) = \frac{1}{2}\left( x^b + y^b\right)^{1/b},
\end{equation}

\noindent where $a$ is a large positive value, and $b$ is a large negative value. For the logarithmic gradient $\nabla$ and the efficiency factor $Q_\text{e}$ we have used $b=-10$. For the stopping time $t_\text{stop}$ we have used $a=30$. 

\section{Compositional Diffusivity}

In this appendix, we discuss the nature and magnitude of the compositional diffusivity,  $\kappa_\mu$, which is an important physical parameter in a double-diffusive thermohaline stability analysis \citep{2013ApJ...768...34B}.

At a minimum, dust diffusivity is provided by its Brownian motion relative to the gas. Following \cite{1996Icar..124..441B}, we estimate the dust compositional diffusivity as
\begin{equation}
\kappa_\text{brown} = \frac{k_\text{B} T}{m} \tau_\text{f},
\end{equation}
where $m$ is the dust grain mass and $\tau_\text{f}$ is the dust friction time.  For a  concrete estimate, we adopt conditions relevant at 1 AU in the protoplanetary disk and assume the Epstein drag regime for micron-sized dust grains is relevant,  leading to $\tau_\text{f} \sim 2.5$s  \citep{2007astro.ph..1485A}.  Using the Epstein formulation for drag,  we can rewrite the compositional diffusivity as

\begin{eqnarray}
\kappa_\text{brown} & \simeq & c_{\text{gas}}   \mu_\text{gas} m_{\text{H}} s^{-2} \rho_{\text{gas}}^{-1} \\ 
                           & \simeq & 0.05 \, \text{cm}^2 \, \text{s}^{-1}  \nonumber
\end{eqnarray}
for a representative sound speed $c_{\text{gas}} = 2.4 \times 10^5 \, \text{cm} \text{s}^{-1}$,  a grain size $s = 10^{-4} \text{cm}$ and a representative gas density $\rho_{\text{gas}} = 5 \times 10^{-10} \, \text{g} \, \text{cm}^{-3}$.  This low-level of diffusivity is further reduced for larger grain sizes and high gas densities. 

Therefore, the diffusivity from Brownian motion evaluates to rather small values of $\kappa_\mu$, typically much less than the kinematic viscosity $ \nu$ for the outer radiative envelopes modeled here (with larger densities found in the gravitationally bound envelopes).

Another potentially relevant process in this context is the possibility that differential settling operates within a compositionally unstable flow. Adopting as a guide the idealized view of nearly-linear thermohaline fingers growing slowly enough that settling happens within them, we might expect the dust to drift with respect to the gas,  so that some dust exchange between the fingers and their environment occurs.  It will likely not act like a diffusion process in modifying dust inhomogeneities, given that the settling is a one-directional process driven by gravity. Nonetheless, we can estimate the magnitude of this effect by computing an effective $\kappa_\mu$ for this dust drift process as the product of the settling velocity with the thermohaline finger size
\begin{equation}
\kappa_\text{settl} \sim v_\text{settl}  d_\text{fing} \simeq \frac{R d_\text{fing}}{T_\text{settl}}.
\end{equation}
This evaluates to $\kappa_\text{settl} \sim 10^{10} \, \text{cm}^2 \, \text{s}^{-1} $  for representative values of $R=10^{11}$~cm,  $T_\text{settl} = 10^{11}$~s and $d_\text{fing} \sim 0.1 \, R$ (see Fig.~\ref{fig:lengthscale}).  We thus obtain values of $\kappa_\mu$ which are considerably larger than those for Brownian motion and potentially well in excess of the gas kinematic viscosity $\nu$. 

Given the large range of possible values for $\kappa_\mu$ suggested by the above estimates, we generalized our thermohaline stability analysis to cases with different values of $\kappa_\mu$, both smaller and larger than the default $\kappa_\mu = \nu$ adopted in the main text.  Reassuringly,  we have recovered nearly identical results to our dispersion relation solutions for the fastest growing mode (\S~3.3), both for $\kappa_\mu << \nu$ and for $\kappa_\mu > \nu$, as long as $\kappa_\mu$ does not exceed $10^3$-$10^4 \times \nu$. In other words, our results are not significantly impacted as long as the condition $ R_0 < 1/\tau$ for thermohaline instability is well satisfied. 

This leads us to conclude that our results are robust to the poorly constrained value of $\kappa_\mu$, as long as it does not exceed the gas kinematic viscosity $\nu$ by several orders of magnitude.  This explains our use of $\kappa_\mu = \nu$ as a reasonable default in the main text.  However, it is clear that further work,  possibly via numerical simulations of the thermohaline instability in a dust-loaded gas, are needed to clarify how efficiently dust inhomogeneities are modified by the process of settling in fully develop compositional turbulence.

\bsp	
\label{lastpage}
\end{document}